\documentclass[a4paper,fleqn,usenatbib]{mnras}
\usepackage{newtxtext,newtxmath}
\usepackage[T1]{fontenc}
\usepackage{ae,aecompl}
\usepackage{graphicx}
\usepackage{pdflscape}

\newcommand{\kms}{\mbox{$\mathrm{km\,s^{-1}}$}}

\newcommand{\MSUN}{\mbox{$\mathrm{M_{\odot}}$}}

\title[Magnetic white dwarf binaries]{Magnetic white dwarfs in post-common-envelope binaries}

\author[S. G. Parsons et al.]{S.~G.~Parsons$^{1}$\thanks{s.g.parsons@sheffield.ac.uk},
B.~T.~G{\"a}nsicke$^{2}$,
M.~R.~Schreiber$^{3,4}$,
T.~R.~Marsh$^{2}$,
R.~P.~Ashley$^{2}$,
\newauthor
E.~Breedt$^{5}$,
S.~P.~Littlefair$^{1}$
and H.~Meusinger$^{6,7}$
\\
$^{1}$ Department of Physics and Astronomy, University of Sheffield,
Sheffield, S3 7RH, UK\\
$^{2}$ Department of Physics, University of Warwick, Coventry CV4 7AL, UK\\
$^{3}$ Departamento de F{\'i}sica, Universidad T{\'e}cnica Santa Mar{\'i}a, Avenida Espa{\~n}a 1680, Valpara{\'i}so, Chile\\
$^{4}$ Millennium Nucleus for Planet Formation (NPF), Valpara{\'i}so, Chile\\
$^{5}$ Institute of Astronomy, University of Cambridge, Madingley Road,
Cambridge CB3 0HA, UK\\
$^{6}$ Th{\"u}ringer Landessternwarte, Sternwarte 5, 07778, Tautenburg, Germany\\
$^{7}$ Universit{\"a}t Leipzig, Fakult{\"a}t f{\"u}r Physik und Geowissenschaften, Linnestra{\ss}e 5, 04103, Leipzig, Germany
}

\date{Accepted 2021 January 27. Received 2021 January 27; in original form 2020 December 17}

\pubyear{2021}

\begin{document}
\label{firstpage}
\pagerange{\pageref{firstpage}--\pageref{lastpage}}
\maketitle

\begin{abstract}

\noindent
Magnitude-limited samples have shown that 20-25 per cent of cataclysmic variables contain white dwarfs with magnetic fields of Mega Gauss strength, in stark contrast to the approximately 5 per cent of single white dwarfs with similar magnetic field strengths. Moreover, the lack of identifiable progenitor systems for magnetic cataclysmic variables leads to considerable challenges when trying to understand how these systems form and evolve. Here we present a sample of six magnetic white dwarfs in detached binaries with low-mass stellar companions where we have constrained the stellar and binary parameters including, for the first time, reliable mass estimates for these magnetic white dwarfs. We find that they are systematically more massive than non-magnetic white dwarfs in detached binaries. These magnetic white dwarfs generally have cooling ages of more than 1\,Gyr and reside in systems that are very close to Roche-lobe filling. Our findings are more consistent with these systems being temporarily detached cataclysmic variables, rather than pre-cataclysmic binaries, but we cannot rule out the latter possibility. We find that these systems can display unusual asymmetric light curves that may offer a way to identify them in larger numbers in future. Seven new candidate magnetic white dwarf systems are also presented, three of which have asymmetric light curves. Finally, we note that several newly identified magnetic systems have archival spectra where there is no clear evidence of magnetism, meaning that these binaries have been previously missed. Nevertheless, there remains a clear lack of younger detached magnetic white dwarf systems.

\end{abstract}


\section{Introduction}

Magnitude-limited surveys have revealed that approximately five per cent of isolated white dwarfs possess megagauss magnetic fields \citep{Kepler13}, generally identified spectroscopically via the Zeeman splitting of their absorption lines. The magnetic incidence is slightly higher in volume limited samples \citep{Tremblay20,McCleery20} and polarimetric observations suggest that almost one in five white dwarfs could possess at least kilogauss magnetic fields \citep{Landstreet19,Bagnulo20}.

A higher magnetic incidence is seen among the white dwarfs in mass transferring cataclysmic variable (CVs). 20-25 per cent of these systems appear to host a magnetic white dwarf \citep{Ferrario15}, but the true incidence could be far higher, with magnetic systems making up 36 per cent of CVs within 150\,pc \citep{Pala20}. However, despite being the direct progenitors of CVs, the incidence of magnetism among detached white dwarf plus main-sequence star binaries is extremely low \citep{Liebert05,Liebert15}, with only a handful of detached magnetic systems serendipitously identified \citep[e.g.][]{Reimers99} among the hundreds of known detached post-common-envelope binaries (PCEBs) \citep{Schreiber10,Rebassa16}. Moreover, all of these detached magnetic white dwarf binaries contain cool ($T_\mathrm{eff} \lesssim 10,000$\,K), hence old, white dwarfs accreting material from the wind of their main-sequence companions and so the progenitors of magnetic CVs remain elusive.

This issue was addressed by \citet{Tout08}, who developed the theory that the magnetic fields of white dwarfs are generated via a dynamo operating during a common-envelope phase. In this scenario all magnetic white dwarfs would have originally been part of binary systems which were close enough that once the progenitor of the white dwarf evolved off the main-sequence it engulfed a companion star. \citet{Tout08} hypothesised that the closer the two stars got within the common-envelope, the stronger the resulting magnetic field of the white dwarf would be, with the highest field strengths resulting from systems that merge during the common envelope, creating isolated, high-field magnetic white dwarfs. This idea also means that those binaries that survive the common-envelope phase with a magnetic white dwarf must be very close and would rapidly evolve into CVs, spending very little time as detached pre-CVs.

However, recently \citet{Belloni20} demonstrated that this model cannot be correct. Using population synthesis they showed that this process would lead to an extremely high magnetic incidence among white dwarfs in close binaries, in fact the overwhelming majority of these binaries should be strongly magnetic. The predicted magnetic field strengths were also too low compared with those measured for white dwarfs in CVs, with this process only able to create white dwarfs with field strengths of around 5\,MG, compared to observations of CVs with typical magnetic field strengths of 7-230\,MG \citep{Ferrario15}. Moreover, this model still produces a substantial number of hot (young) magnetic white dwarfs in close detached binaries, a population that is completely absent among the observed samples. It therefore remains unclear how magnetic CVs are created and why there is an absence of young magnetic white dwarfs in detached pre-CVs.

The small number of old magnetic white dwarfs in detached binaries represent a crucial population for understanding how magnetic CVs are formed and evolve. If they are genuinely pre-CV systems (i.e. they have been detached since the end of the common envelope phase) then this would lend support to the idea that magnetic CVs descent from detached magnetic PCEBs and these systems would be pre-polars (PREPs, \citealt{Schwope09}). On the other hand, if they are magnetic CVs that have temporarily detached (so-called low accretion rate polars, LARPs, \citealt{Schwope02}), then this would imply a complete lack of magnetic white dwarfs in pre-CVs, meaning that the magnetic fields of the white dwarfs in CVs would need to be generated during the CV phase. However, two factors have prevented a detailed investigation of this population, firstly, only ten systems are currently known \citep{Ferrario15} which limits any conclusions that can be drawn since the sample is subject to heavy selection effects. Secondly, the stellar and binary parameters of these systems are poorly constrained, since these binaries are single-lined and the spectra often heavily contaminated by strong cyclotron emission, making direct measurements difficult. Precise measurements of the fundamental parameters of these binaries, such as the white dwarf masses and Roche-lobe filling factors, would be a powerful result that could distinguish between the pre-CV or temporarily detached CV scenarios. 

In this paper we identify four new magnetic white dwarfs in detached binaries. We analyse these binaries in detail, as well as two previously identified detached magnetic white dwarf binaries, to precisely determine the stellar and binary parameters for all six systems. We then discuss what these measurements imply for the evolutionary history of these systems.

\section{Target selection}

The small number of previously discovered magnetic white dwarfs in detached binaries with low mass stellar companions were predominantly identified via cyclotron emission lines in their optical spectra, which mark them out from the otherwise similar spectra of non-magnetic white dwarfs in detached PCEBs. While these cyclotron lines allow a confident classification of a system as possessing a magnetic white dwarf, it is likely that not all magnetic systems exhibit strong cyclotron lines at optical wavelengths, as the strength of the cyclotron lines is related to the accretion rate of material onto the white dwarf. In these detached binaries it is thought that the accretion occurs via the white dwarf capturing material from the wind of the main-sequence star. This makes the strength of the emission lines sensitive to the binary separation. A wider orbit will decrease the accretion rate, making the cyclotron lines weaker and possibly undetectable. Moreover, depending upon the orientation of the binary and the location of the magnetic poles on the white dwarf, the strength of the cyclotron lines can vary substantially over the binary orbit, potentially even disappearing entirely if the magnetic poles rotate out of view \citep[e.g.][]{Reimers99,Schmidt05_2}. Finally, most known magnetic white dwarfs in detached binaries have field strengths of several tens of megagauss, where the cyclotron lines appear in the optical. However, lower magnetic field strengths will shift these lines towards longer wavelengths, meaning that signs of magnetism may no longer be evident in the optical spectrum. This is the case for the eclipsing magnetic white dwarf in SDSS\,J030308.35+005444.1, which possesses a magnetic field with a strength of only 8\,MG \citep{Parsons13_mag}. In this case the cyclotron lines appear at infrared wavelengths and the optical spectrum shows no clear indication of magnetism \citep{Pyrzas09}. Therefore, it is possible to miss magnetic white dwarf systems if only a single low resolution optical spectrum is available.

While the optical spectrum of SDSS\,J030308.35+005444.1 shows no cyclotron lines, it does hint that there is something unusual about the white dwarf. The temperature of this white dwarf is thought to be around 9\,000\,K \citep{Pyrzas09,Parsons13_mag}. A non-magnetic white dwarf of this temperature should possess strong Balmer absorption lines that would be easily detectable in the Sloan Digital Sky Survey (SDSS) spectrum of this source, even after accounting for the dilution from the M dwarf companion. However, there are no obvious Balmer lines in the SDSS spectrum and the white dwarf was classified as a featureless DC white dwarf \citep{Rebassa16}. High resolution data revealed that the lack of Balmer absorption lines is due to a combination of Zeeman splitting and additional Balmer emission from the white dwarf \citep{Parsons13_mag}, hence magnetism is clearly the cause of the lack of Balmer absorption lines in the SDSS spectrum of this system. Apart from cyclotron emission lines, very few white dwarfs in these systems show (Zeeman split) Balmer absorption lines. Therefore, it is worth investigating other systems previously classified as DC+dM binaries for signs of magnetism.

On this premise, the sample presented in this paper covers the majority of systems classified as close DC+dM binaries in the SDSS white dwarf main-sequence binary catalogue of \citet{Rebassa16}. Our sample also contains two previously known, but poorly studied, detached magnetic white dwarf binaries (IL\,Leo and HS\,0922+1333), which were included in order to understand the general properties of this population. We also included two systems with unusual Catalina Real Time Transient Survey (CRTS) light curves (SDSS\,J0750+4943 and SDSS\,J2229+1853), where magnetism was suspected as the cause of the highly asymmetric shape of the light curves \citep{Parsons15}. Since the majority of our observations were obtained in the southern hemisphere, we were unable to observe northern DC+dM binaries in the \citet{Rebassa16} catalogue, therefore our sample is not a complete study of all known DC+dM binaries. In total this paper presents new spectroscopic data for 11 systems.

Finally, seven new candidate detached magnetic white dwarf binaries were discovered during a search for unusual quasars in Kohonen maps of SDSS spectra \citep{Meusinger12,Meusinger16}. These systems are discussed in Section~\ref{sec:new_cands}.

\section{Observations and their reduction}

A full list of our observations is given in Table~\ref{tab:obslog}.

\begin{table}
 \centering
  \caption{Journal of spectroscopic observations. For X-shooter observations we list the number of spectra obtained and exposure times in the UVB, VIS and NIR arms.}
  \label{tab:obslog}
  \tabcolsep=0.05cm
  \begin{tabular}{@{}lccccc@{}}
  \hline
  Telescope/ & Date & No. of  & Exopsure & Transparency & seeing \\
  Instrument &      & spectra & times (s) &             & (arcsec) \\
  \hline
\multicolumn{3}{l}{\bf SDSS J022503.02+005456.2:}\\
VLT/X-shooter & 31 Aug 2017 & 5/8/13 & 600/350/240 & Clear & 0.5 \\
VLT/X-shooter & 16 Sep 2017 & 5/8/13 & 600/350/240 & Clear & 0.6 \\
\multicolumn{3}{l}{\bf SDSS J075015.11+494333.2:} \\
INT/IDS & 11 Feb 2015 & 13 & 900 & Thin clouds & 1.0 \\
\multicolumn{3}{l}{\bf SDSS J084841.17+232051.7:} \\
VLT/X-shooter & 15 Apr 2017 & 6/11/31 & 600/300/120 & Clear & 0.5 \\
VLT/X-shooter & 18 Apr 2018 & 15/27/75 & 600/300/120 & Clear & 0.9 \\
\multicolumn{3}{l}{\bf SDSS J085336.03+072033.5:} \\
VLT/X-shooter & 17 Apr 2017 & 10/18/30 & 600/300/200 & Clear & 0.8 \\
\multicolumn{3}{l}{\bf HS 0922+1333:} \\
VLT/X-shooter & 15 Apr 2017 & 10/18/30 & 600/300/200 & Clear & 0.5 \\
\multicolumn{3}{l}{\bf IL Leo:} \\
VLT/X-shooter & 16 Apr 2017 & 20/18/22 & 300/334/300 & Clear & 0.6 \\
\multicolumn{3}{l}{\bf SDSS J114030.06+154231.5:} \\
VLT/X-shooter & 16 Apr 2017 & 10/18/30 & 600/300/200 & Clear & 0.6 \\
VLT/X-shooter & 17 Apr 2017 & 5/9/15 & 600/300/200 & Clear & 0.8 \\
VLT/X-shooter & 15 May 2018 & 5/9/15 & 600/300/200 & Clear & 0.5 \\
\multicolumn{3}{l}{\bf SDSS J131632.04-003758.0:} \\
VLT/X-shooter & 16 Apr 2017 & 5/9/15 & 600/300/200 & Clear & 0.6 \\
\multicolumn{3}{l}{\bf SDSS J145238.12+204511.9:} \\
VLT/X-shooter & 16 Apr 2017 & 10/18/30 & 600/300/200 & Clear & 0.6 \\
\multicolumn{3}{l}{\bf SDSS J220848.32+003704.6:} \\
VLT/X-shooter & 30 Aug 2017 & 5/8/13 & 600/350/240 & Thin clouds & 0.7 \\
VLT/X-shooter & 4 Sep 2017 & 5/8/13 & 600/350/240 & Clear & 0.6 \\
\multicolumn{3}{l}{\bf SDSS J222918.95+185340.2:} \\
INT/IDS & 25 Jul 2015 & 3 & 900 & Clear & 1.2 \\
VLT/X-shooter & 30 Aug 2017 & 10/18/44 & 600/300/135 & Thin clouds & 0.6 \\
VLT/X-shooter & 31 Aug 2017 & 10/18/44 & 600/300/135 & Clear & 0.5 \\
  \hline
\end{tabular}
\end{table}

\subsection{VLT/X-shooter spectroscopy}

The bulk of our spectroscopic observations were performed with the medium resolution echelle spectrograph X-shooter \citep{Vernet11}, which is mounted at the Cassegrain focus of the VLT-UT2 at Paranal, Chile. X-shooter covers the spectral range from the atmospheric cutoff in the UV to the near-infrared K band in three separate arms, known as the UVB (0.30$-$0.56 microns), VIS (0.56$-$1.01 microns) and NIR (1.01$-$2.40 microns). Separate slit widths can be set for each arm and our observations were performed with slit widths of 1.0, 0.9 and 0.9 arcsec in the UVB, VIS and NIR arms respectively. We also binned the detector in the VIS arm by a factor of 2 in the spatial direction, while the UVB arm was binned by a factor of 2 in both the spatial and dispersion directions This results in a resolution of R$\sim$5000 in the UVB and NIR arms and R$\simeq$8000 in the VIS arm. All of the data were reduced using the standard X-shooter pipeline release (version 3.5.0) within {\sc esoreflex}.

The VIS arm spectra were telluric corrected using {\sc molecfit} \citep{Smette15,Kausch15}. Since our observations were performed in stare mode (in order to minimise overheads in the optical) the NIR spectra have poor sky subtraction and therefore no telluric correction was attempted for these data. The fit to the telluric features was also used to correct for small (typically $\sim$1{\kms}) systemic velocity offsets in the data via the method described in \citet{parsons17}. All spectra were then placed on a heliocentric wavelength scale.

\subsection{INT/IDS spectroscopy}

SDSS\,J0750+4943 and SDSS\,J2229+1853 were both observed with the Intermediate Dispersion Spectrograph (IDS) mounted on the Cassegrain focus of the 2.5m Isaac Newton Telescope (INT) at La Palma, Spain. Both objects were observed using the R300V grating centred at 6460{\AA} and a slit width of 1{\arcsec}, resulting in a dispersion of 1.9{\AA}/pixel over a wavelength range of 4000-8700{\AA}.

The data were reduced using the {\sc pamela}\footnote{{\sc pamela} is available through the {\sc starlink} software package \citep{Currie14}} and {\sc molly}\footnote{{\sc molly} is available from \url{http://cygnus.astro.warwick.ac.uk/phsaap/software/}} packages. The spectra were first bias subtracted and flat-fielded, then optimally extracted. The wavelength calibration was performed using CuAr+CuNe lamps exposed regularly throughout each night. The observations of spectral standard stars were used to flux calibrate our spectra, but no telluric correction was applied to the IDS spectra. 

\section{Spectroscopic analysis}

\begin{figure*}
  \begin{center}
    \includegraphics[width=\textwidth]{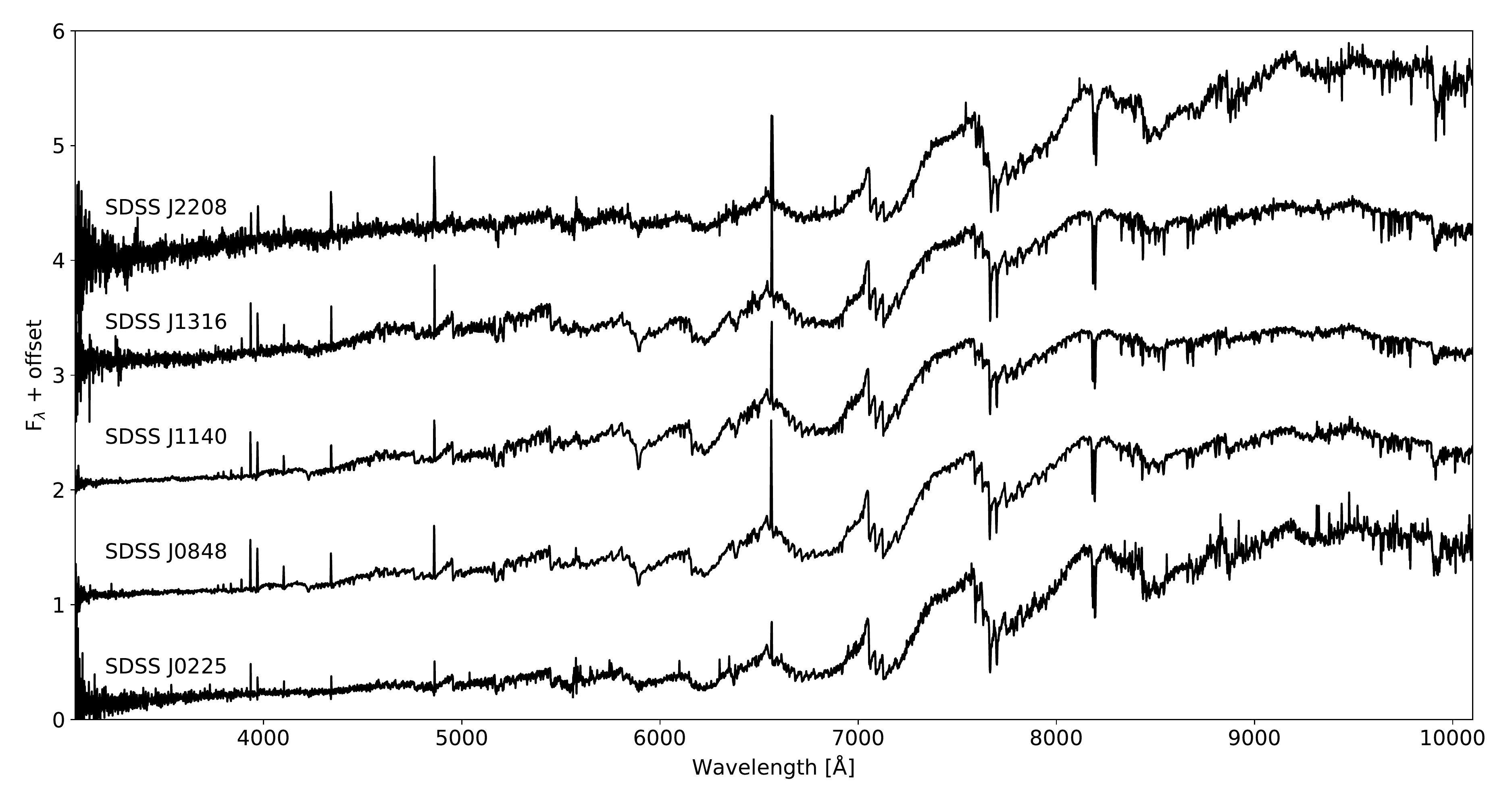}
    \caption{Normalised averaged spectra of the non-magnetic white dwarf plus M dwarf binaries analysed in this paper. The spectra are completely dominated by the M dwarf components, the white dwarf is only evident from a small excess at the shortest wavelengths. Emission lines from the M dwarf are seen in all systems indicating that these are active stars (since irradiation effects are negligible in these systems).}
  \label{fig:DC_specs}
  \end{center}
\end{figure*}

\begin{figure*}
  \begin{center}
    \includegraphics[width=\textwidth]{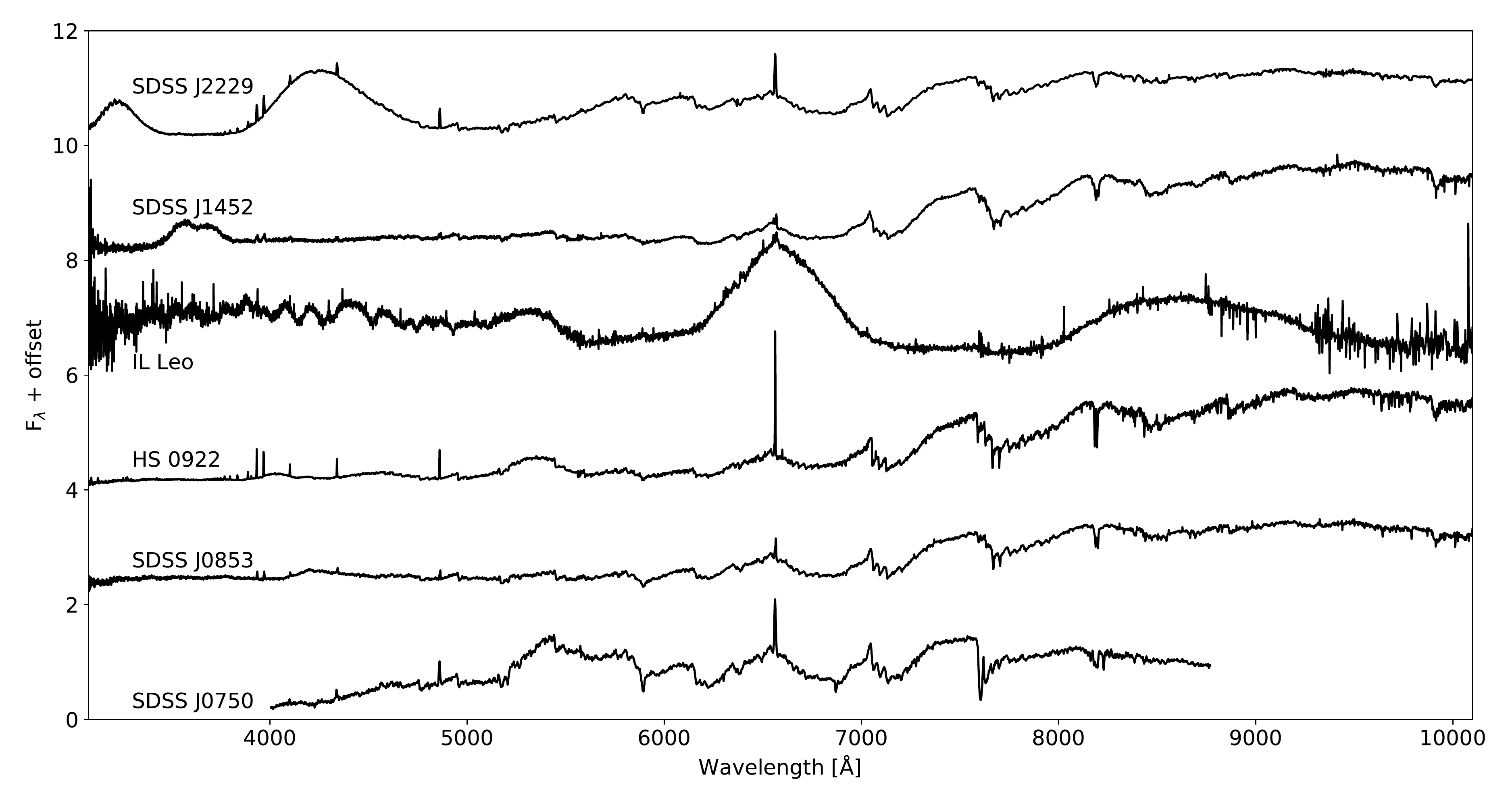}
    \caption{Normalised averaged spectra of the magnetic white dwarf plus M dwarf binaries analysed in this paper. The M dwarf components are visible in all the systems apart from IL\,Leo. Cyclotron emission lines are seen in all systems, although the strength of these lines varies substantially among systems. Zeeman-split hydrogen Balmer absorption lines from the white dwarf are evident in the spectrum of IL\,Leo.}
  \label{fig:MAG_specs}
  \end{center}
\end{figure*}

The sample of systems analysed in this paper can be split into two groups, those containing a magnetic white dwarf and those with a non-magnetic white dwarf. Unless otherwise stated, the spectra of the M dwarfs in all the systems presented in this paper (both magnetic and non-magnetic) were analysed identically. The normalised averaged spectra for all of the binaries analysed in this paper are shown in Figures~\ref{fig:DC_specs} and \ref{fig:MAG_specs} for the non-magnetic and magnetic systems respectively.

\subsection{Radial velocity measurements}

For all but one system analysed in this paper there are no clean features from the white dwarf to help constrain its radial velocity, the exception being SDSS\,J0848+2320, which contains a DZ white dwarf with clear calcium H/K absorption lines (this white dwarf was originally classified as a DC white dwarf based on the SDSS spectrum but our higher resolution data revealed that it is actually a DZ). Therefore, in order to constrain the binary and physical parameters, radial velocity measurements were obtained for only the M dwarf components in these systems, except for SDSS\,J0848+2320, where we used features from both stars.

Radial velocities were measured in each spectrum by fitting the sodium absorption doublet from the M dwarf at $\simeq$8200{\AA}. The lines were fitted with a combination of a straight line and two Gaussian components. In order to improve accuracy, we fitted all spectra of the same system simultaneously, keeping the width and depth of the Gaussian components the same for all spectra and only allowed the velocity of the Gaussians to vary between spectra (note that we specifically targeted the sodium absorption doublet because it is free from any cyclotron emission in all of the magnetic systems). The best fitting values and their uncertainties were found using the Levenberg-Marquardt method \citep{Press86} and are listed in Table~\ref{tab:velocities}. The same method was used to fit the Ca\,{\sc ii} absorption line from the white dwarf in SDSS\,J0848+2320, although in this case a single Gaussian component was used. These values are listed in Table~{\ref{tab:wd_velocities}.

Orbital periods, radial velocity semi-amplitudes of the main-sequence stars and systemic velocities were determined by fitting the velocity measurements with a constant plus sine wave over a range of periods and computing the $\chi^2$ of the resulting fit. We also included any archival velocity measurements from \citet{Rebassa16}. The orbital periods for most of the systems analysed in this paper have been previously measured, but our additional high-precision velocity measurements greatly improve the precision of the ephemerides as well as radial velocity semi-amplitude measurements. These updated values are listed in Table~\ref{tab:params}. Radial velocity curves are plotted in Figure~\ref{fig:RV_plots}.

\begin{figure*}
  \begin{center}
    \includegraphics[width=0.965\textwidth]{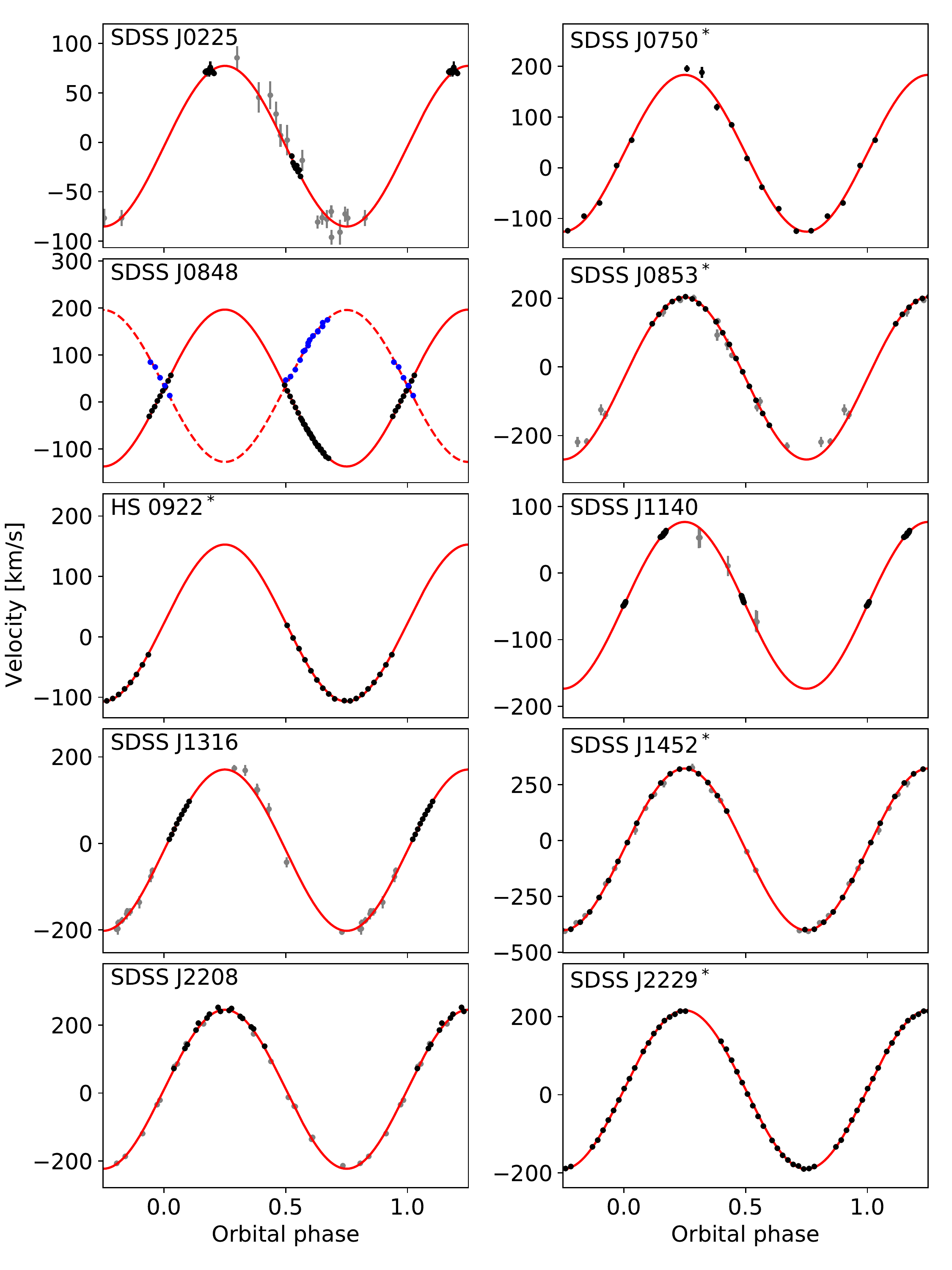}
    \caption{Radial velocity measurements and fits (red lines). Black points are new measurements for the M dwarfs in these systems, previous measurements are shown in grey (taken from \citealt{Rebassa16}). Systems containing magnetic white dwarfs are highlighted with an asterisk ($^*$). For SDSS\,J0848+2320 we also show the velocity measurements for the white dwarf in blue and the fit to these (red, dashed line).}
  \label{fig:RV_plots}
  \end{center}
\end{figure*}

\subsubsection{Radial velocity measurements for IL\,Leo}

\begin{figure*}
  \begin{center}
    \includegraphics[width=0.97\textwidth]{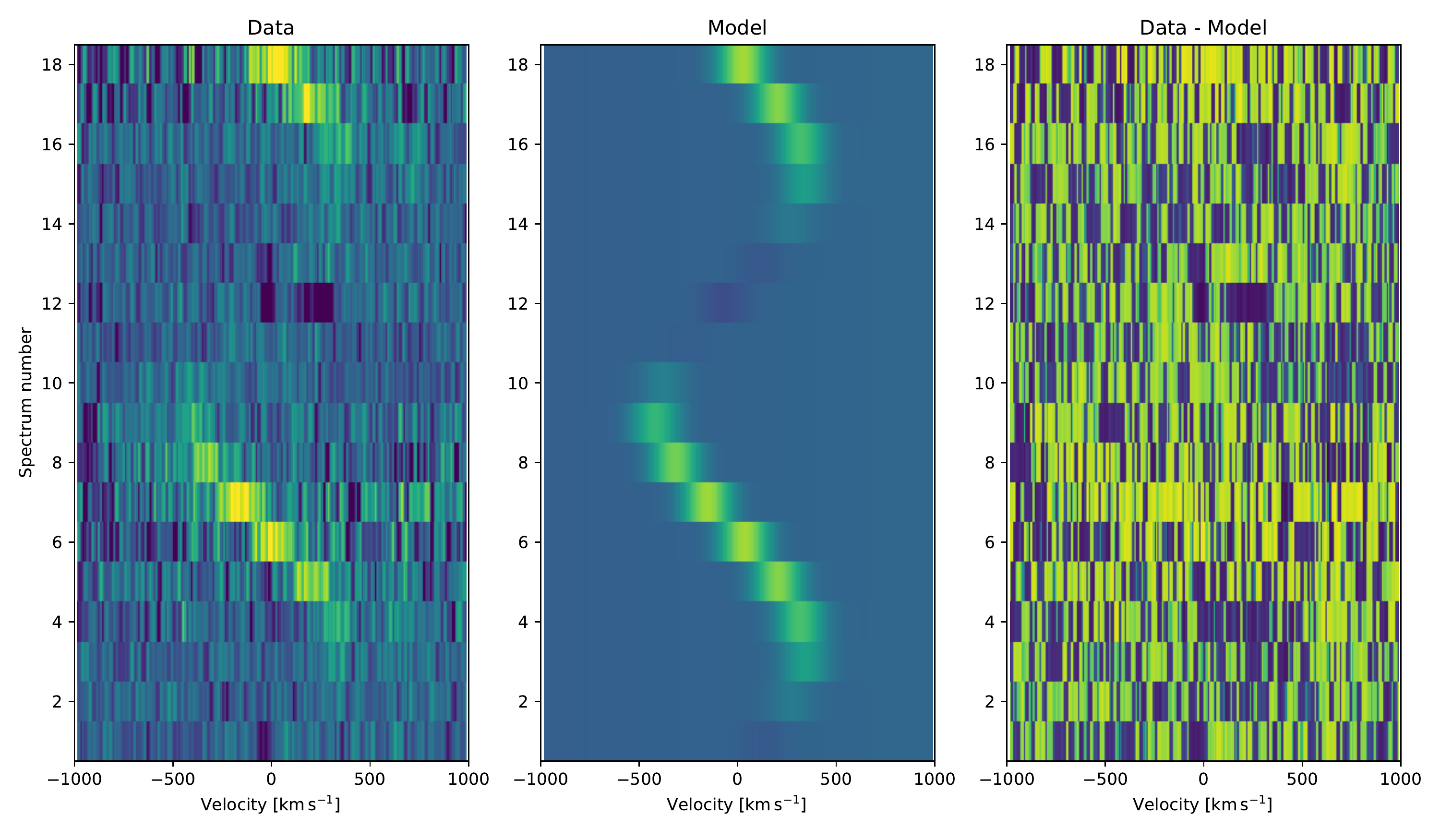}
    \caption{Trailed spectra of the H$\alpha$ emission line in the magnetic system IL\,Leo with time running upwards. The left-hand panel shows the VLT/Xshooter spectra. The centre panel shows our best fit model to the irradiation-driven emission line from the M dwarf component. The right-hand panel shows the residuals of the fit.}
  \label{fig:ILLeo_RV}
  \end{center}
\end{figure*}

There are no absorption features from the M dwarf in the magnetic system IL\,Leo, likely due to its faintness and the strong cyclotron lines in this system (see Figure~\ref{fig:MAG_specs}). Therefore, a different method was needed to constrain the physical and binary parameters of this system. The only feature visible from the M dwarf in this system is a H$\alpha$ emission line, which varies in strength over the orbit, peaking in strength at an orbital phase of 0.5, implying that it is caused by irradiation from the white dwarf. A reliable fit to this emission line in individual spectra is difficult due to the weakness of this line. Therefore, we decided to simultaneously fit all the spectra of IL\,Leo around the H$\alpha$ line with a model in which the position of the emission line changes sinusoidally according to $\gamma + K_\mathrm{em}\sin{(2\pi\phi)}$, where $\gamma$ is the systemic velocity, $K_\mathrm{em}$ is the radial velocity semi-amplitude of the emission line and $\phi$ is the orbital phase. The strength of the line also varies according to ($1-\cos{\phi})/2$ to simulate the behaviour of the line due to irradiation (see \citealt{Parsons17_K2} for a detailed description of the model). The best fit model is shown in Figure~\ref{fig:ILLeo_RV} and yields a radial velocity semi-amplitude of $K_\mathrm{em}=384.2\pm4.5$\,\kms. However, since this line is irradiation-driven, the emission is concentrated on the inner hemisphere of the M dwarf. As such, the velocity of this emission line does not track the centre-of-mass of the star and so this is only a lower limit on the true radial velocity semi-amplitude of the star.

IL\,Leo is the only system to show clear Balmer absorption lines from the white dwarf in its spectrum (see Figure~\ref{fig:MAG_specs}). However, these lines can not be used to measure the radial velocity of the white dwarf due to a combination of the low signal-to-noise ratio of the spectra at these wavelengths (the lines are only apparent when all the spectra of this system are combined) as well as the fact that the lines are affected by the magnetic field of the white dwarf. The lines are clearly Zeeman split and, since the field strength varies over the surface of the white dwarf, the amount each component is shifted will vary throughout the orbit, as different regions of the white dwarf rotate into view, making any velocity measurements unreliable.

\subsection{Rotational broadening measurements}

The rotational broadening of the M dwarf in these close binaries is sensitive to the binary and stellar parameters and thus it can be used to place useful constraints on the stellar masses for example (see Section~\ref{sec:params}). Therefore, we tried to measure the rotational broadening in all our systems. Unfortunately there are no absorption features from the M dwarf in IL\,Leo so it was not possible to measure the rotational broadening in this system. Moreover, SDSS\,J0750+4943 only has data from the INT/IDS. The resolution of these data is insufficient to measure a reliable rotational broadening, so this system also lacks a measurement. Rotational broadening measurements were performed for all the other systems we observed.

The rotational broadening was measured by artificially broadening the lines of template stars to fit the observed line profiles of our systems, taking into account any additional smearing of the lines from the velocity shift of the main-sequence star during an exposure. Template M dwarf spectra were taken from \citet{parsons17}. All of these template stars were observed with X-shooter using an identical instrumental setup to the binaries presented in this paper. Note that we applied a high-pass filter to both the observed and broadened template spectra before comparing them in order to prevent the continuum dominating the rotational broadening determination. We fitted just the wavelength range around the sodium 8200\,{\AA} absorption doublet, since this wavelength range is free from any cyclotron emission in all the magnetic systems. The best fit rotational broadening measurements for our systems are listed in Table~\ref{tab:params}. We reached a precision of better than 2.5\,{\kms} in all our systems, although in the case of the longer period systems (where the rotational broadening is much smaller) this precision still results in a large fractional uncertainty. 

\section{Stellar and binary parameters} \label{sec:params}

In this section we summarise the methods used to constrain the stellar and binary parameters in both the magnetic and non-magnetic systems. All stellar and binary parameters for our systems are given in Table~\ref{tab:params}, along with literature values for similar systems that have been studied previously.

\subsection{M dwarf masses and radii} \label{sec:md_mr}

We estimated the masses of the M dwarfs in all our systems using the mass-luminosity relation for late-type stars from \citet{Mann19}. This relates the masses of M dwarfs to their absolute $K_S$-band magnitudes. All of our systems have good Gaia early DR3 parallaxes \citep{Gaia_eDR3} and 2MASS $K_S$-band magnitudes \citep{2MASS}, with the exception of IL\,Leo, which lacks a $K_S$-band measurement due to its faintness. Moreover, IL\,Leo is the only system in which the M dwarf does not overwhelmingly dominate the $K_S$-band flux. As such, even if the system had a $K_S$-band measurement the mass-luminosity relation would be difficult to use in this case since the white dwarf contribution would need to be subtracted beforehand. Therefore, the parameters of IL\,Leo were constrained using a different technique that is summarised in Section~\ref{sec:IL_Leo}. For all other systems the $K_S$-band flux is completely dominated by the M dwarf. The field strengths of the white dwarfs in all the magnetic systems we observed are at least 60\,MG (except IL\,Leo) and thus the cyclotron emission will be limited to wavelengths shorter than 1.8 microns ($H$ band).

The mass-luminosity relation from \citet{Mann19} yields masses with statistical uncertainties of 2-3 per cent. We used the distance estimates to our systems from \citet{Bailer18} and propagated the uncertainties in both the distances and 2MASS $K_S$-band magnitudes when calculating the M dwarf masses. Typical uncertainties in the M dwarf masses are $\sim$5 per cent. However, we note that the \citet{Mann19} relation is based on M dwarfs in wide, astrometric binaries and may not be as suitable for the rapidly rotating and Roche-distorted M dwarfs in the close systems studied in this paper. Indeed \citet{Parsons18} found that this relation tended to slightly underestimate the masses of M dwarfs in close binaries with white dwarfs possibly due to the enhanced number of starspots expected on the tidally locked and hence rapidly rotating M dwarfs in close binaries. \citet{Parsons18} found that the masses estimated from the mass-luminosity relation could be up to 5 per cent underestimated. We therefore set a lower limit of 5 per cent on the uncertainties of the M dwarf masses in this paper to account for this systematic uncertainty. The effects of metallicity are expected to be negligible for values in the solar neighbourhood \citep{Mann19}, hence we assume solar metallicity for all our targets.
 
We estimated the radii of the M dwarfs using a semi-empirical mass-radius relationship for M dwarfs, based on a combination of a large dataset of M dwarf radius measurements \citep{Morrell19} and the mass-luminosity relation from \citet{Mann19}. This semi-empirical mass-radius relationship will be presented in more detail in a future publication (Brown et~al., in prep) and leads to more accurate radius predictions compared to theoretical mass-radius relationships, which tend to under-predict the radii of low-mass stars by 5-10 per cent \citep{Morales05,Parsons18}. Any effects from metallicity are likely to be at a level smaller than the 5 per cent systematic error we placed on the M dwarf masses (which are propagated to the radius estimates).

\subsection{White dwarf masses} \label{sec:wd_mass}

The mass of the white dwarfs in these binary systems can be constrained using the rotational broadening measurements ($V_\mathrm{rot}\sin{i}$) via
\begin{equation}
V_\mathrm{rot}\sin{i} = K_\mathrm{MS}(1+q)\frac{R_\mathrm{MS}}{a}, \label{eqn:vsini}
\end{equation}
where $K_\mathrm{MS}$ is the radial velocity semi-amplitude of the M dwarf, $q=M_\mathrm{WD}/M_\mathrm{MS}$, the binary mass ratio, $R_\mathrm{MS}$ is the radius of the M dwarf and $a$ is the semi-major axis \citep{Marsh94}. This relation is only valid if the M dwarf is synchronously rotating, which is expected to be the case in these extremely close binary systems.

\begin{landscape}
\begin{table}
 \centering
  \caption{Stellar and binary parameters for PCEBs containing cool white dwarfs. Systems analysed in this paper are highlighted in italic. M dwarf masses and Roche-lobe filling factors for the systems analysed in this paper were determined assuming that the star is in thermal equilibrium, hence the uncertainties on these values are purely statistical. T0 corresponds to the inferior conjunction of the main-sequence star. $G$ is the {\it Gaia} magnitude, which (along with the parallax measurements) are taken from early DR3. RLFF is the Roche-lobe filling factor of the M dwarf component. References: (1) \citet{Parsons12_ucool}, (2) this paper, (3) \citet{Zorotovic16}, (4) \citet{Parsons13_mag}, (5) \citet{Schmidt05_2}, (6) \citet{Reimers00}, (7) \citet{Tovmassian07}, (8) \citet{Reimers99}, (9) \citet{Schwarz01}, (10) \citet{Schmidt07}, (11) \citet{Schwope09}, (12) \citet{Schmidt05}, (13) \citet{Burleigh06}, (14) \citet{Farihi08}, (15) \citet{Breedt12}, (16) \citet{Szkody03}, (17) \citet{Southworth15}, (18) \citet{Szkody08}, (19) \citet{Kafka10}. }
  \label{tab:params}
  \tabcolsep=0.05cm
  \begin{tabular}{@{}lcccccccccccccc@{}}
  \hline
  Name & $G$ & $\pi$ & P$_\mathrm{orb}$ &  T0 & K$_\mathrm{MS}$ & $\gamma_\mathrm{MS}$ & $v\sin{i}$ & Spectral & M$_\mathrm{MS}$ & M$_\mathrm{WD}$ & T$_\mathrm{eff,WD}$ & $B$\footnotemark[1] & RLFF & Reference \\
  & (mag) & (mas) & (days) & (HJD) & ($\kms$) & ($\kms$) & ($\kms$) & type & (\MSUN) & (\MSUN) & (K) & (MG) & \\
  \hline
  \multicolumn{4}{l}{\bf DC/Z+dM binaries:} \\
SDSS\,J0138-0016   & 16.43 & $20.36\pm0.07$ & 0.07276491(2) & 2455867.507405(6) & $346.7\pm2.3$ & $84.5\pm1.2$ & - & M5.0 & $0.132\pm0.003$ & $0.53\pm0.01$ & $3570\pm100$ & - & $0.91\pm0.03$ & (1) \\
{\it SDSS\,J0225+0054}   & 18.99 & $4.49\pm0.27$ & 0.9210733(2) & 2456407.7947(51) & $81.2\pm2.0$ & $-4.0\pm1.3$ & $5.5\pm2.3$ & M4.5 & $0.166\pm0.015$ & $0.55\pm0.40$ & $6600\pm300$ & - & $0.20\pm0.02$ & (2) \\
{\it SDSS\,J0848+2320}   & 17.05 & $3.61\pm0.10$ & 0.3717195(1) & 2458043.67163(5) & $167.0\pm0.5$ & $29.5\pm0.5$ & $69.5\pm2.5$ & M3.0 & $0.430\pm0.022$ & $0.44\pm0.02$ & $7600\pm500$ & - & $0.58\pm0.02$ & (2) \\
{\it SDSS\,J1140+1542}   & 16.90 & $3.03\pm0.11$ & 3.11329(3) & 2455807.491(19) & $125.2\pm0.9$ & $-48.5\pm0.5$ & $7.3\pm1.5$ & M2.5 & $0.475\pm0.024$ & $>$1.22 & $8900\pm900$ & - & $0.15\pm0.01$ & (2) \\
{\it SDSS\,J1316$-$0037} & 17.66 & $5.42\pm0.14$ & 0.4027340(2) &  2456241.50583(95) & $186.5\pm1.1$ & $-15.5\pm2.0$ & $26.5\pm2.0$ & M4.0 &  $0.202\pm0.010$ & $0.64\pm0.11$ & $7300\pm200$ & - & $0.38\pm0.02$ & (2) \\
{\it SDSS\,J2208+0037}   & 18.93 & $4.45\pm0.34$ & 0.10337209(1) & 2457588.67776(4) & $233.9\pm0.6$ & $11.3\pm0.5$ & $68.2\pm2.2$ & M5.0 & $0.150\pm0.032$ & $0.46\pm0.24$ & $6100\pm400$ & - & $0.81\pm0.17$ & (2,3) \\
  \multicolumn{4}{l}{\bf Magnetic WD+dM binaries:} \\
SDSS\,J0303+0054 & 17.36 & $8.30\pm0.10$ & 0.1344376668(1) & 2453991.617307(2) & $339.9\pm0.3$ & $14.9\pm0.2$ & $81.7\pm1.1$ & M4.5 & 0.181-0.205 & 0.825-0.853 & $\sim$9150 & 8 & $0.84\pm0.03$ & (4) \\
{\it SDSS\,J0750+4943} & 16.08 & $4.66\pm0.05$ & 0.17300847(5) & 2457065.49219(70) & $154.6\pm3.9$ & $28.5\pm4.3$ & - & M2.5 & $0.460\pm0.023$ & $0.94\pm0.12$ & $15000_{-1000}^{+1400}$ & 98 or 196 & $0.98\pm0.02$ & (2) \\
SDSS\,J0837+3830\footnotemark[2] & 18.86 & $0.98\pm0.23$ & 0.1325 & - & - & - & - & M5.0 & $0.447\pm0.083$ & - & - & 65 & $>$0.97 & (5) \\
{\it SDSS\,J0853+0720} & 18.02 & $3.74\pm0.15$ & 0.15021618(2) & 2456138.68614(20) & $236.3\pm0.6$ & $-33.2\pm0.5$ & $62.4\pm2.3$ & M4.0 & $0.221\pm0.016$ & $0.83\pm0.15$ & $9000\pm300$ & 84 & $0.82\pm0.05$ & (2) \\
{\it HS\,0922+1333} & 16.46 & $6.06\pm0.09$ & 0.1683125(5) & 2457859.61706(10) & $129.9\pm2.3$ & $22.8\pm2.4$ & $58.6\pm2.2$ & M3.5 & $0.366\pm0.018$ & $0.71\pm0.07$ & $<$8500 & 66(81) & $0.95\pm0.03$ & (2,6,7) \\
WX\,LMi & 16.34 & $10.31\pm0.05$ & 0.1159229(13) & 2451157.6439(9) & - & - & - & M4.5 & $0.225\pm0.003$ & - & - & 60(68) & $>$0.91 & (8,9) \\
{\it IL\,Leo}\footnotemark[3] & 19.30 & $1.93\pm0.23$ & 0.05709(18) & 2457860.61634(15) & $>$ $384.2\pm4.5$ & $-43.2\pm5.0$ & - & $>$M6 & $<$0.090 & $>$0.48 & $<$11000 & 42 & $>$0.95 & (2,10) \\
SDSS\,J1059+2727 & 20.30 & $1.30\pm0.94$ & $>$0.125 & - & - & - & - & M4.0 & - & - & $<$8500 & 57 & - & (10) \\
SDSS\,J1206+5100 & 18.64 & $3.30\pm0.16$ & 0.1576726(6) & 2458254.7030(5) & - & - & - & M3.5 & $0.208\pm0.023$ & - & $\sim$9000 & 108 or 216 & $>$0.727 & (11) \\
SDSS\,J1212+0136\footnotemark[2]$^{,}$\footnotemark[3] & 17.98 & $6.47\pm0.13$ & 0.0614081(7) & 2453686.5276(1) & $>$ $355\pm6$ & - & - & L5 & $<$0.080 & $>$0.41 & $\sim$9500 & 7 & $>$0.90 & (12,13,14) \\
SDSS\,J1250+1549\footnotemark[3] & 18.20 & $7.58\pm0.14$ & 0.05995(1) & 2455646.7056(1) & $>$ $360.9\pm2.5$ & $12.8\pm2.1$ & - & M8 & $<$0.085 & $>$0.42 & $\sim$10000 & 20 & $>$0.90 & (15) \\
PZ\,Vir & 20.48 & $1.78\pm1.64$ & 0.11022(7) & - & - & - & - & M6.0 & - & - & - & 64 & - & (15,16) \\
{\it SDSS\,J1452+2045} & 17.89 & $7.97\pm0.12$ & 0.106265437(9) & 2456207.8708(1) & $362.0\pm0.4$ & $-39.5\pm0.5$ & $82.4\pm1.5$ & M5.0 & $0.150\pm0.008$ & $0.83\pm0.08$ & $<$6500 & 300 & $0.84\pm0.03$ & (2,3) \\
SDSS\,J1514+0744\footnotemark[3] & 18.67 & $5.53\pm0.18$ & 0.061610(1) & 2455646.89093(3) & $>$ $362.8\pm0.7$ & $-28.0\pm0.8$ & - & L3 & $<$0.085 & $>$0.44 & $\sim$10000 & 36 & $>$0.90 & (15) \\
MQ\,Dra & 17.34 & $5.51\pm0.06$ & 0.182985(5) & 2454156.9138(1) & - & - & - & M5.0 & $0.238\pm0.008$ & - & $\sim$8000 & 60 & $>$0.70 & (16,18)\\
SDSS\,J2048+0050\footnotemark[2] & 18.27 & $1.66\pm0.16$ & 0.175 & - & $110.4\pm3.9$ & $-36.8\pm1.5$ & - & M3.0 & $0.446\pm0.044$ & - & - & 62 & $>$0.96 & (5,19) \\
{\it SDSS\,J2229+1853} & 16.51 & $3.44\pm0.06$ & 0.1891844(1) & 2457996.12995(3) & $203.1\pm0.2$ & $12.9\pm0.5$ & $105.5\pm1.2$ & M3.0 &  $0.460\pm0.023$ & $0.76\pm0.05$ & $<$8500 & 84 & $0.96\pm0.03$ & (2) \\
  \hline
\end{tabular}
\end{table}
\footnotetext[1]{Parentheses denote secondary pole}
\footnotetext[2]{System shows some evidence of (episodic) Roche-lobe overflow}
\footnotetext[3]{Potential period bounce cataclysmic variable}
\end{landscape}

With a measurement of the rotational broadening, radial velocity semi-amplitude, mass and radius of the M dwarf then Equation~\ref{eqn:vsini} can be used to solve for the mass of the white dwarf, since $a$ is only dependent upon the orbital period ($P$) and stellar masses,
\begin{equation}
a^3 = \frac{G(M_\mathrm{WD} + M_\mathrm{MS})P^2}{4 \pi^2},
\end{equation}
thus combining these two equations yields the white dwarf mass. Uncertainties in all the parameters were propagated through these equations to determine the statistical error on the white dwarf mass. The final error on the white dwarf mass is strongly dependent on how precisely the rotational broadening can be measured. For systems where the M dwarf is more slowly rotating this results in substantial uncertainty on the white dwarf mass (e.g. for SDSS\,J0225+0054, which has a period of 21.7 hours).

We used the stellar masses, orbital period and M dwarf radius to estimate the Roche-lobe filling factor (RLFF) for each of our systems. These filling factors (listed in Table~\ref{tab:params}) are given as the ratio of the radius of the M dwarf to its volume-averaged Roche-lobe radius, as opposed to the linear RLFF which is given by the distance from the centre of mass of the M dwarf to its surface in the direction of the L1 point divided by the distance to the L1 point. Uncertainties in the stellar and binary parameters were propagated to determine the error on the RLFF.

\subsection{White dwarf temperatures}

\begin{figure*}
  \begin{center}
    \includegraphics[width=0.495\textwidth]{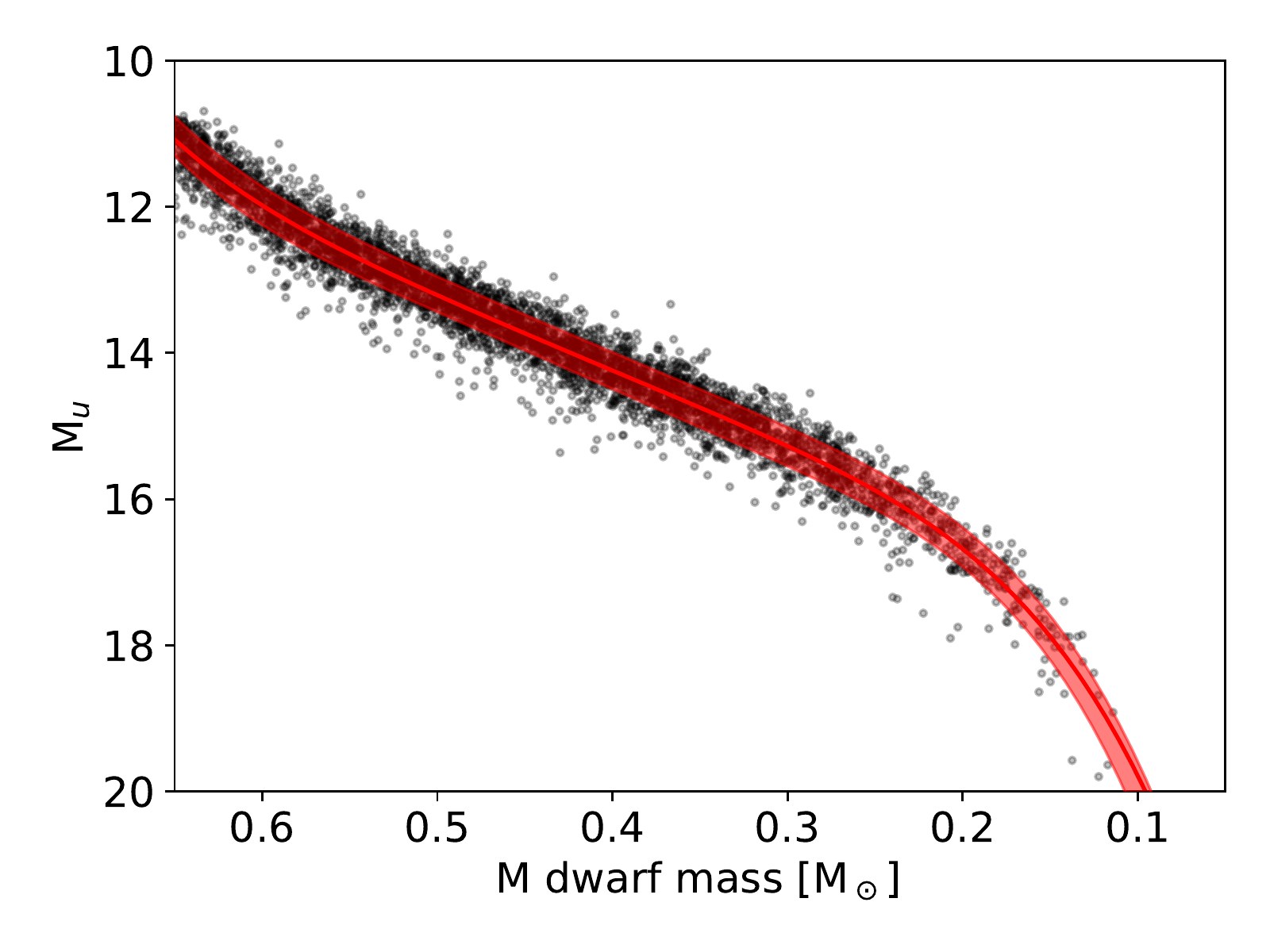}
    \includegraphics[width=0.495\textwidth]{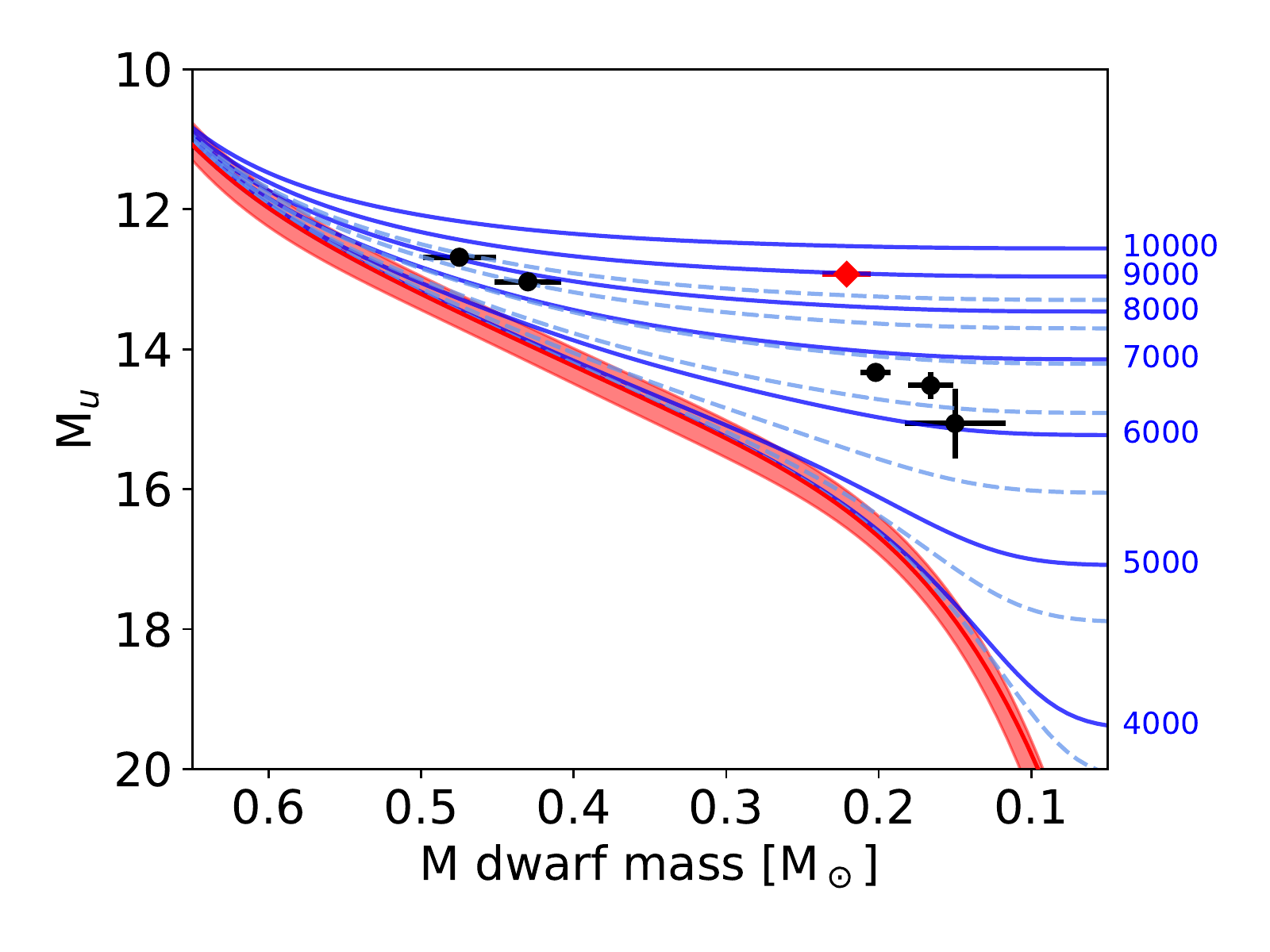}
    \caption{{\it Left:} The mass-luminosity relation for M dwarf stars using SDSS $u$-band magnitudes. The sample of M dwarfs was taken from \citet{Morrell19}. Masses were determined using the $K_S$-band mass-luminosity relation from \citet{Mann19}. A fifth order polynomial fit is shown in red. The shaded region marks the 16th to 84th percentile region of the measurements. {\it Right:} The change in the $u$-band absolute magnitude of an M dwarf when a white dwarf is placed next to it. Solid blue lines indicate white dwarfs with different temperatures (labelled on the right) and surface gravities of $\log{g}=8.0$, while the dashed blue lines are for white dwarfs with $\log{g}=8.5$. The black points mark the values for the non-magnetic systems. The magnetic system SDSS\,J0853+0720 is indicated by the red diamond.}
  \label{fig:absU}
  \end{center}
\end{figure*}

\begin{figure*}
  \begin{center}
    \includegraphics[width=0.92\textwidth]{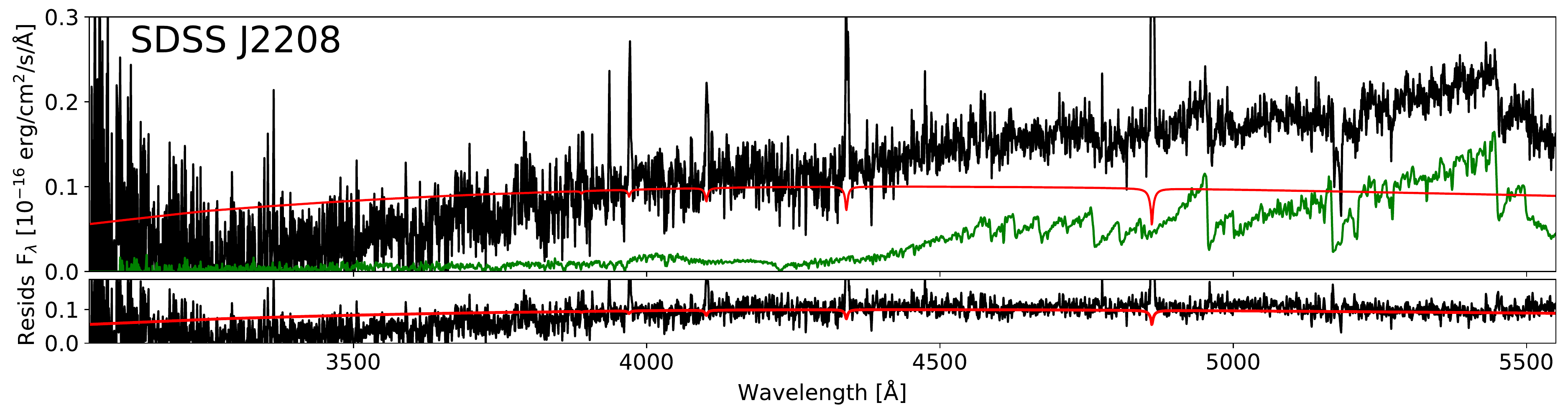}
    \includegraphics[width=0.92\textwidth]{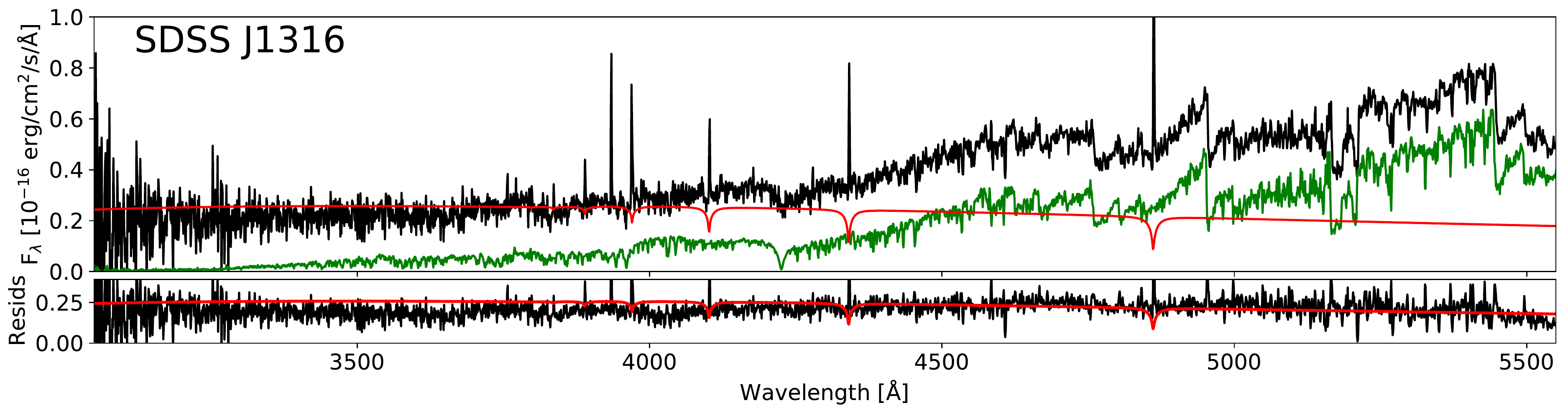}
    \includegraphics[width=0.92\textwidth]{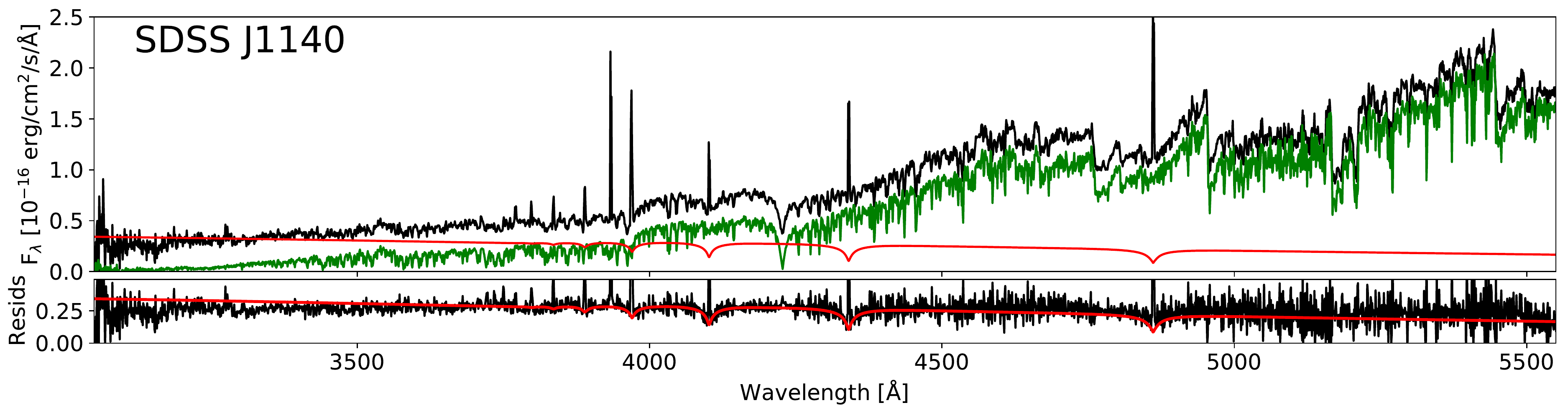}
    \includegraphics[width=0.92\textwidth]{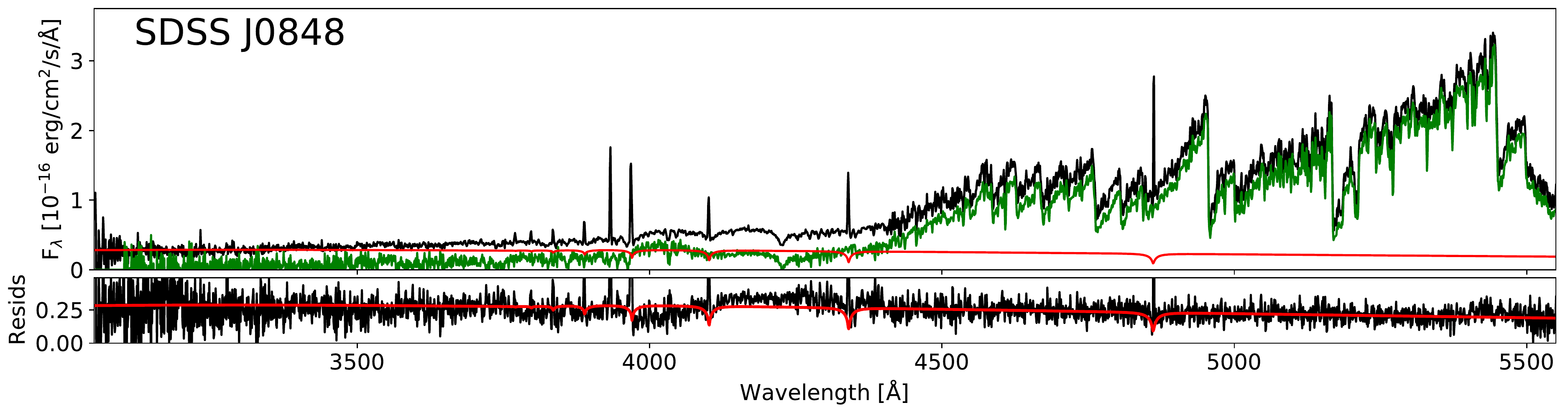}
    \includegraphics[width=0.92\textwidth]{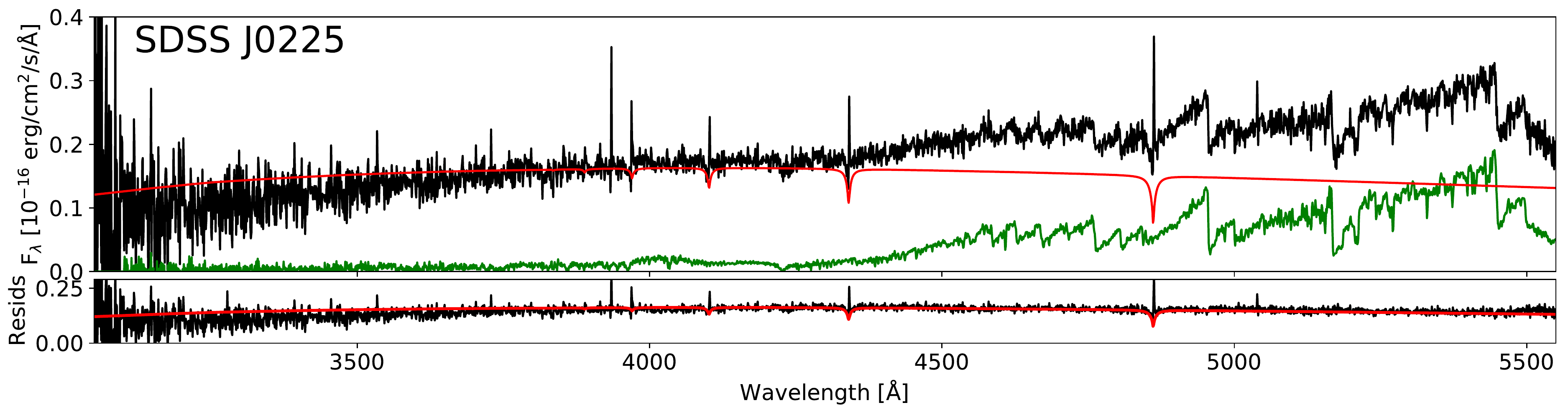}
    \caption{Average spectra (black lines) of the non-magnetic systems in our sample. For each system the best fit M dwarf template spectrum is shown in green and a white dwarf model with the temperature determined from the relation shown in Figure~\ref{fig:absU} and scaled to the distance of the system is shown in red. The M dwarf subtracted spectrum is plotted below each spectrum with the scaled white dwarf model over-plotted. The white dwarf models were not fitted to the spectroscopic data, but the good agreement between the models and the data supports the reliability of the method used to determine the white dwarf temperatures. The overestimation of the white dwarf models at the shortest wavelengths is likely due to slit losses in the X-shooter data.}
  \label{fig:spec_fits}
  \end{center}
\end{figure*}

The temperature of the white dwarf is a useful measurement to make, since it helps constrain the age of the white dwarf and thus the system. However, this is a difficult measurement to make because we specifically selected systems containing featureless DC white dwarfs, since we suspected that these could be hiding magnetic fields. This means that none of the white dwarfs in our sample have clear white dwarf features that can be fitted to constrain their temperature. Moreover, the vast majority of our systems lack any ultraviolet data, preventing us from constraining the white dwarf temperatures at shorter wavelengths. The situation is made even harder for the magnetic systems, since the cyclotron emission dilutes and distorts any underlying emission from the photosphere of the white dwarf.

The white dwarfs do contribute a small but detectable amount of optical flux in these systems. Most of these systems were originally identified through excess blue flux in their SDSS spectra \citep{Rebassa16} and so we can use this to constrain the white dwarf temperatures. However, the M dwarf contribution must be accounted for if accurate white dwarf temperatures are to be obtained.

We constructed a mass-luminosity relation for M dwarf stars using SDSS $u$-band data. The advantage of using the $u$-band is that there are relatively few features in the spectra of M dwarfs in this wavelength range. This mitigates the substantial scatter seen in the $V$-band luminosities of M dwarfs caused by the large metallicity spread of field M dwarfs \citep{Delfosse00}. The disadvantage of using the $u$-band being the intrinsic faintness of low-mass stars at these wavelengths. Fortunately the sample of M dwarfs presented by \citet{Morrell19} contains many bright, nearby stars. Cross-matching this catalogue with SDSS gives 4564 M dwarfs with reliable $u$-band magnitudes and parallaxes. Since all of these M dwarfs also have $K_S$-band magnitudes we computed their masses using the \citet{Mann19} mass-luminosity relation and then constructed our own mass-luminosity relation from the $u$-band data, which is shown in the left-hand panel of Figure~\ref{fig:absU}. The amount of scatter in this relation is similar to the $K_S$-band mass-luminosity relation \citep{Mann19}, although there are relatively few measurements for stars below 0.15\,{\MSUN} in the $u$-band. A fifth order polynomial fit to this relation gives
\begin{eqnarray}
M_u & = & (26.52\pm0.09) - (106.61\pm1.29) M_* \nonumber \\
& & + (450.63\pm6.86) M_*^2 - (1058.07\pm17.50) M_*^3  \nonumber \\
& & + (1255.07\pm21.50) M_*^4 - (603.48\pm10.23)  M_*^5
\end{eqnarray}
where $M_*$ is the mass of the M dwarf, valid between 0.1\,{\MSUN} and 0.65\,{\MSUN}.

We used this relation to estimate the M dwarf contribution to the $u$-band fluxes in all the non-magnetic systems. The excess $u$-band flux was then determined to be from the white dwarf. Since both the distance and white dwarf mass are already known, the only other parameter that affects the $u$-band luminosity of a white dwarf is its temperature, assuming a standard mass-radius relationship \citep{Fontaine01} and a hydrogen atmosphere. The effect of adding a white dwarf with a varying temperature is shown in the right-hand panel of Figure~\ref{fig:absU}, demonstrating that this is a particularly useful technique for measuring the temperatures of very cool white dwarfs in binaries with low-mass stars. In general this technique allows us to constrain the white dwarf temperature to better than 500\,K, although the uncertainty rises as the M dwarf mass increases and this approach is not ideal for systems with M dwarfs more massive than around 0.5\,{\MSUN}.

We tested the reliability of this method by scaling a white dwarf model \citep{Koester10} to the distance of each system with the parameters determined from the $u$-band fit and comparing this model to the X-shooter spectrum of each system. The result of this is shown in Figure~\ref{fig:spec_fits}. For each system we plot part of the X-shooter spectrum along with the best fit M dwarf template spectrum (in green) and the scaled white dwarf model (in red). We also show the M dwarf subtracted spectra with white dwarf models over-plotted. In each case the predicted white dwarf flux matches the residual flux after the M dwarf is subtracted, demonstrating that the white dwarf parameters estimated from the $u$-band flux are reliable. Note that due to slit losses the white dwarf models tend to over-predict the flux at the shortest wavelengths. This imperfect calibration is why we have not attempted to constrain the white dwarf temperatures from the X-shooter data directly.

We were able to use the same approach to measure the temperature of the magnetic white dwarf in SDSS\,J0853+0720 because the SDSS data for this system were obtained when the magnetic pole was not visible, although we note that the tracks in the right-hand panel of Figure~\ref{fig:absU} are for non-magnetic white dwarfs, which may not be appropriate for the magnetic white dwarf in SDSS\,J0853+0720. The use of this method was not possible on any of the other magnetic systems. In these cases we used the spectra taken when the cyclotron lines were at their weakest and subtracted off the M dwarf component (by scaling the best fit template spectrum used in the rotational broadening measurements). We then scaled white dwarf model spectra \citep{Koester10} to the distance of the system, keeping the mass fixed at the measured value and using a mass-radius relationship \citep{Fontaine01}. We varied the temperature of the white dwarf until the flux was clearly over predicted between the cyclotron lines (see Figure~\ref{fig:mag_temps} for example) and used this to set an upper limit on the white dwarf temperature. However, since the white dwarf models we used are for non-magnetic white dwarfs and the M dwarf components are never perfectly removed, the temperature estimate for all the magnetic systems should be taken with some caution.

\begin{figure}
  \begin{center}
    \includegraphics[width=\columnwidth]{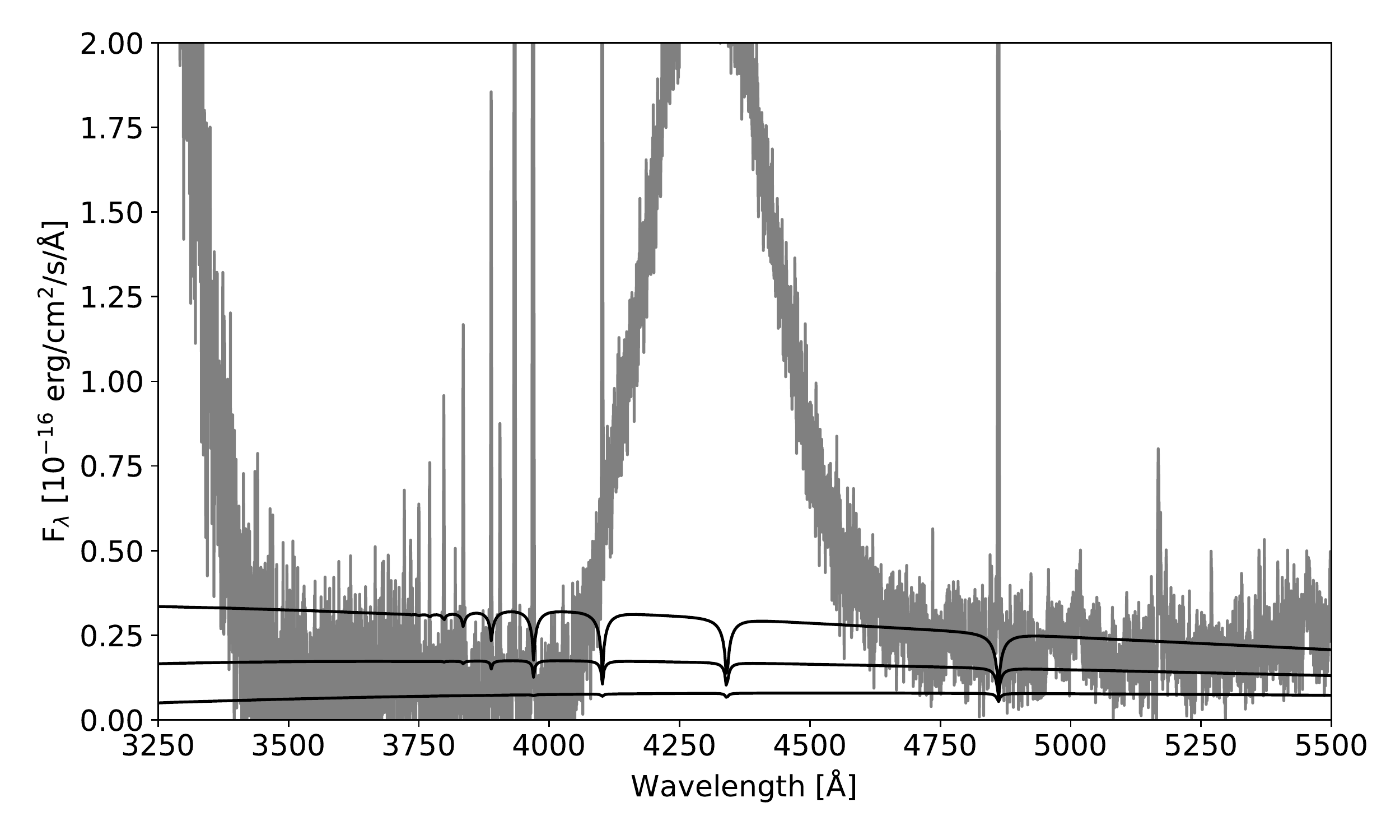}
    \caption{Spectrum of the magnetic system SDSS\,J2229+1853 taken at minimum light (when the cyclotron lines are at their weakest) with the M dwarf component removed. Overplotted are non-magnetic white dwarf models with varying temperatures (8000\,K, 7000\,K and 6000\,K) and the same mass as measured for the white dwarf, scaled to the distance of the system (assuming a standard mass-radius relationship, \citealt{Fontaine01}). We place an upper limit on the white dwarf temperature, above which the continuum flux (between cyclotron lines) would be overestimated.}
  \label{fig:mag_temps}
  \end{center}
\end{figure}

\begin{figure*}
  \begin{center}
    \includegraphics[width=0.87\textwidth]{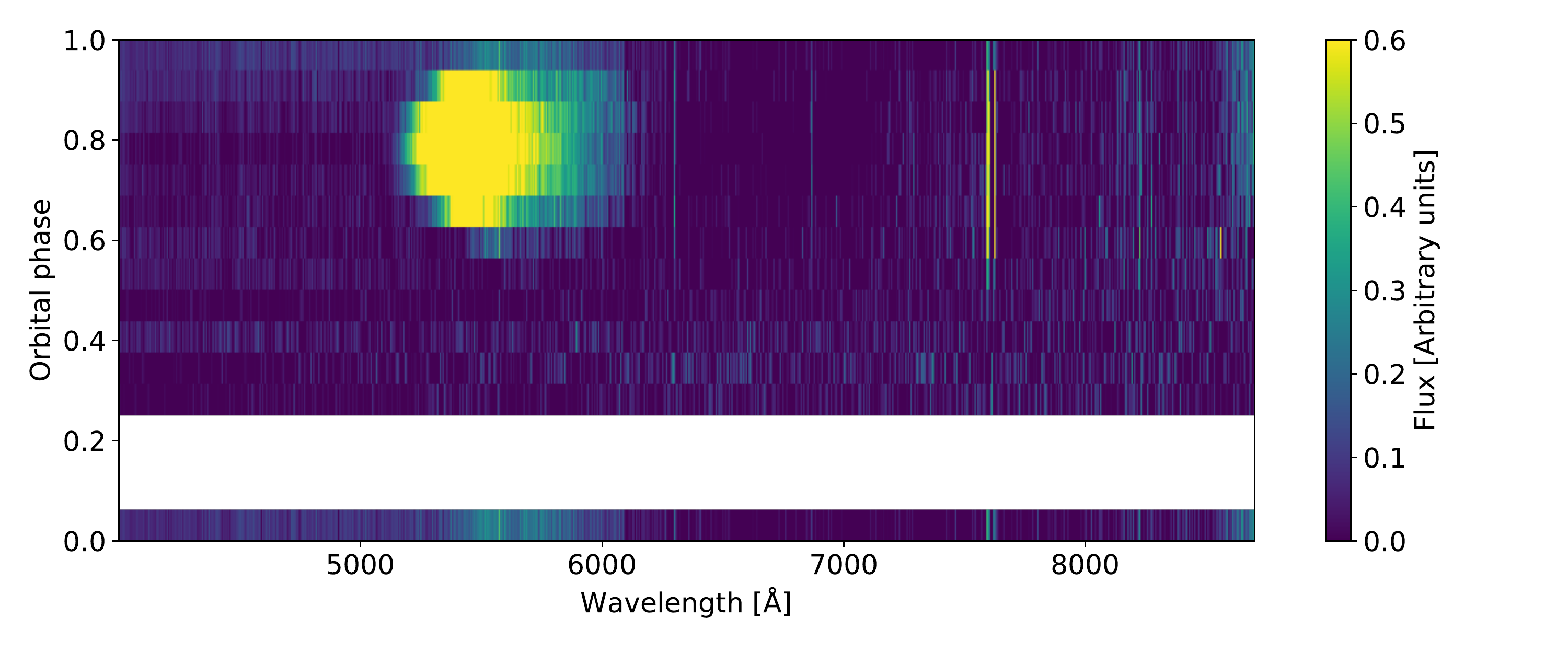}
    \caption{Phase-folded trailed spectra of SDSS\,J0750+4943 with the M dwarf component removed.}
  \label{fig:SDSS0750_trail}
  \end{center}
\end{figure*}

\begin{figure*}
  \begin{center}
    \includegraphics[width=0.87\textwidth]{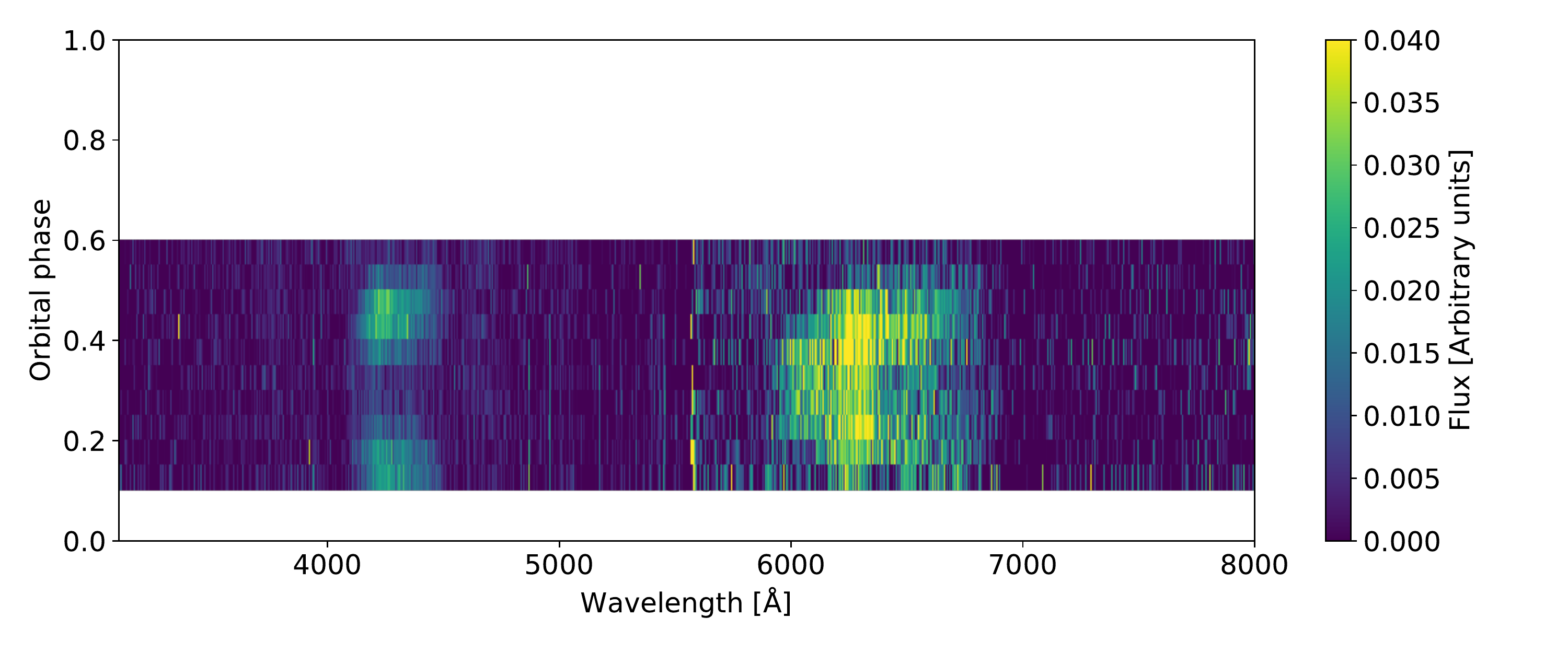}
    \caption{Phase-folded trailed spectra of SDSS\,J0853+0720 with the M dwarf component removed.}
  \label{fig:SDSS0853_trail}
  \end{center}
\end{figure*}

\begin{figure*}
  \begin{center}
    \includegraphics[width=0.87\textwidth]{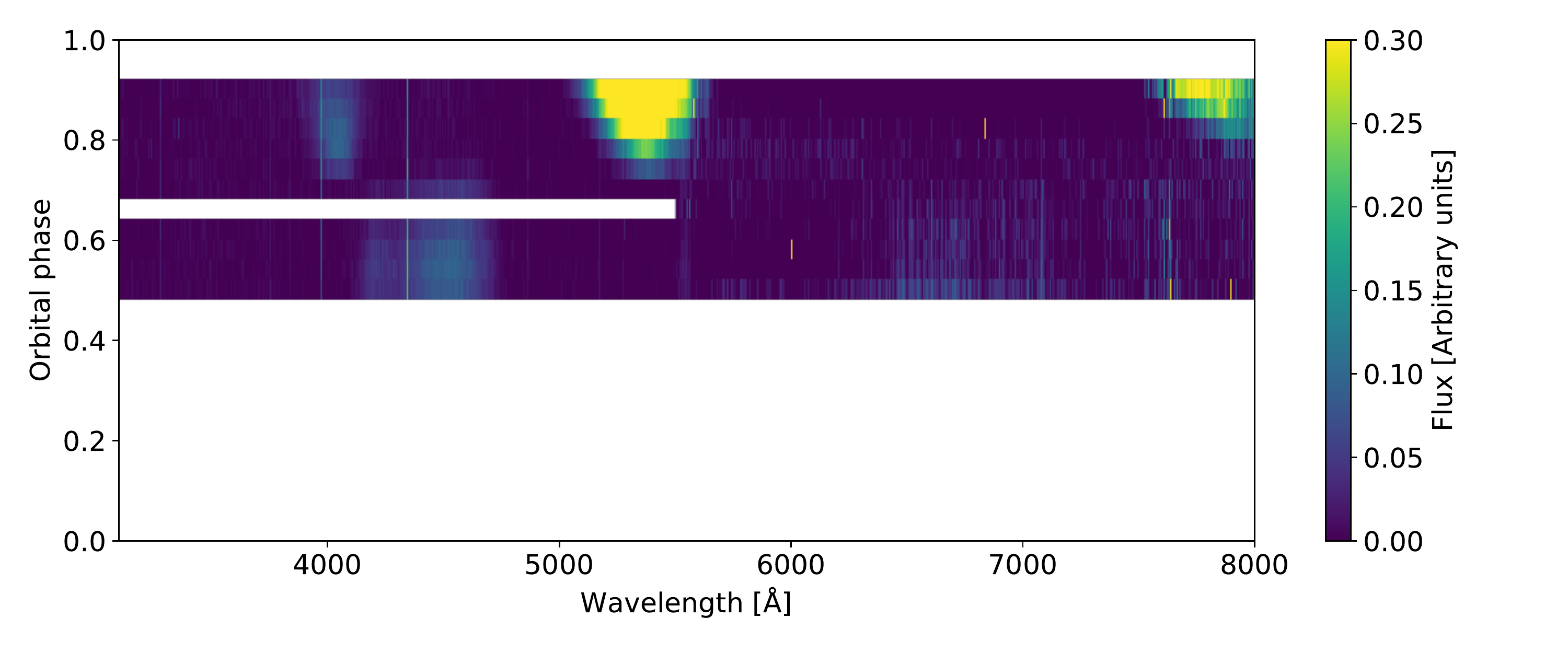}
    \caption{Phase-folded trailed spectra of HS\,0922+1333 with the M dwarf component removed.}
  \label{fig:HS0922_trail}
  \end{center}
\end{figure*}

\begin{figure*}
  \begin{center}
    \includegraphics[width=0.87\textwidth]{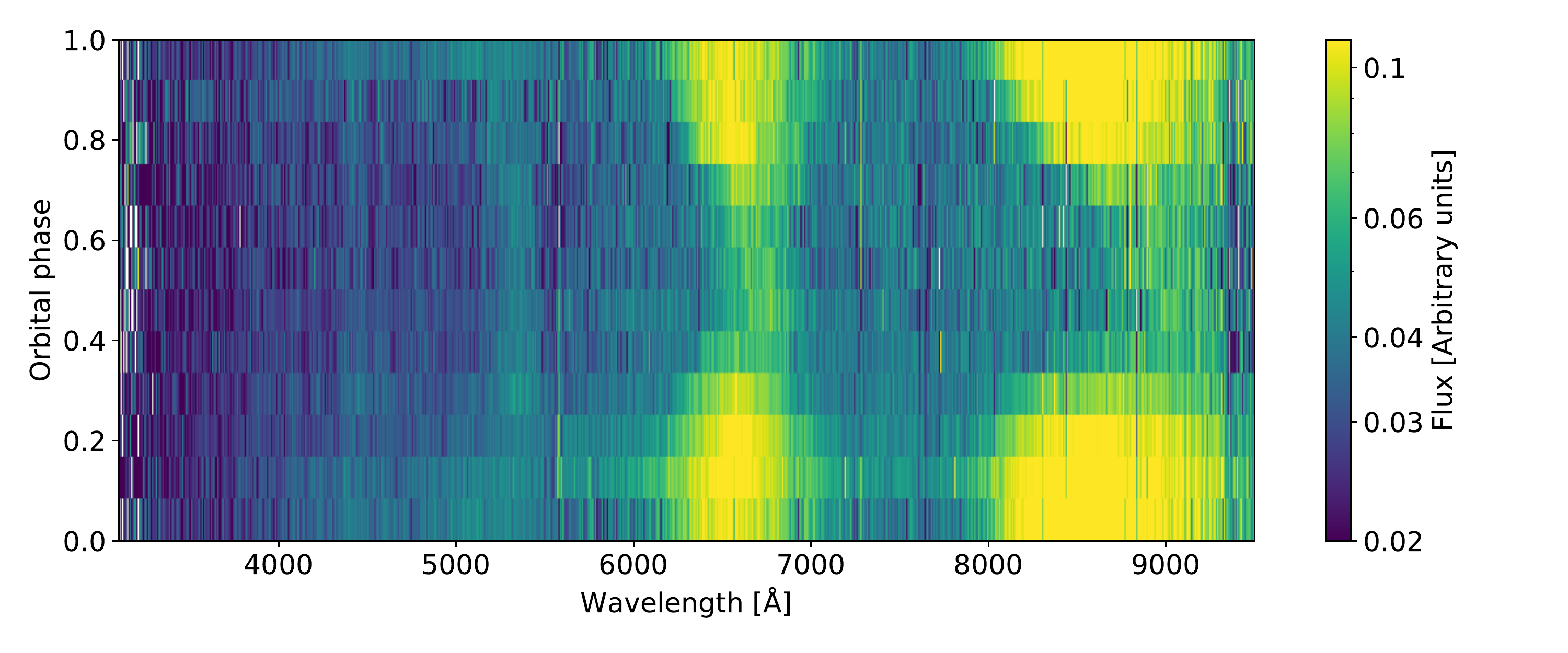}
    \caption{Phase-folded trailed spectra of IL\,Leo. Since the M dwarf component is so weak in the optical there is no need to remove it in this case.}
  \label{fig:ILLeo_trail}
  \end{center}
\end{figure*}

\begin{figure*}
  \begin{center}
    \includegraphics[width=0.87\textwidth]{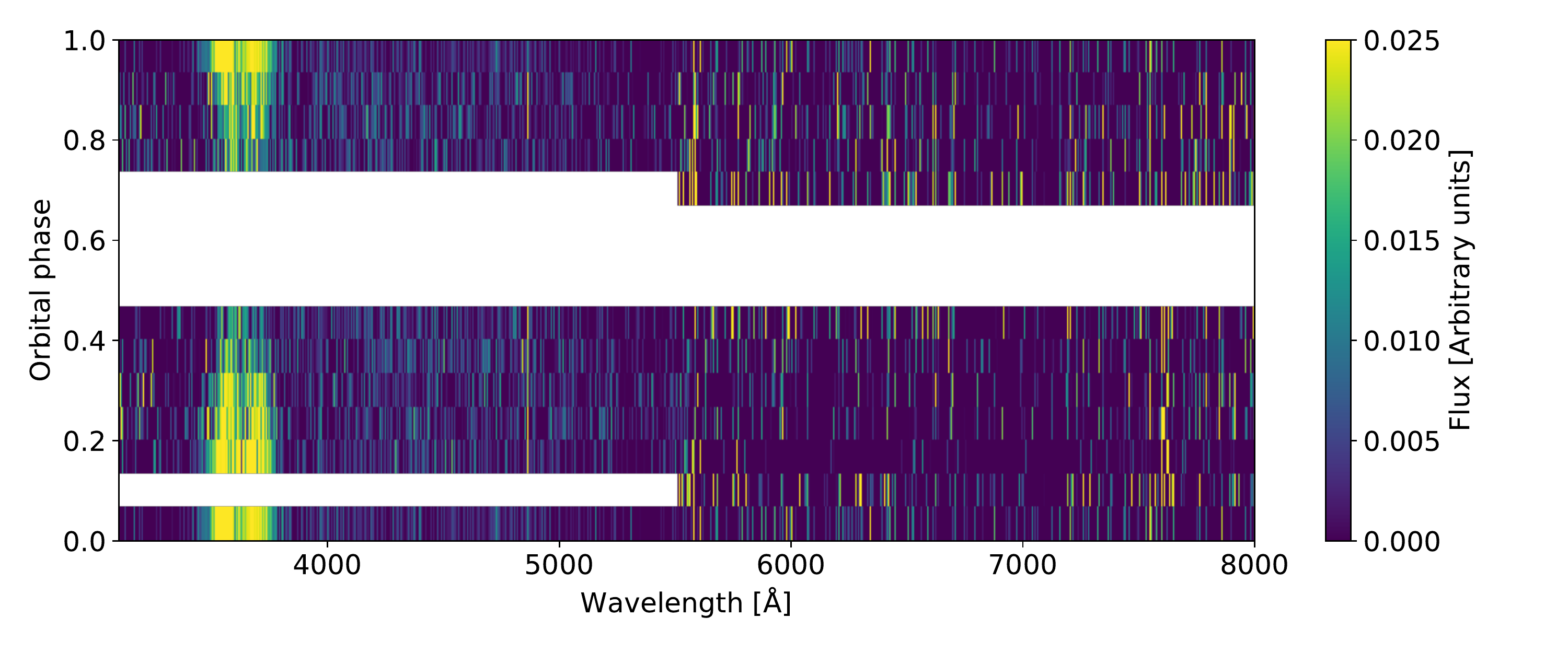}
    \caption{Phase-folded trailed spectra of SDSS\,J1452+2045 with the M dwarf component removed.}
  \label{fig:SDSS1452_trail}
  \end{center}
\end{figure*}

\begin{figure*}
  \begin{center}
    \includegraphics[width=0.87\textwidth]{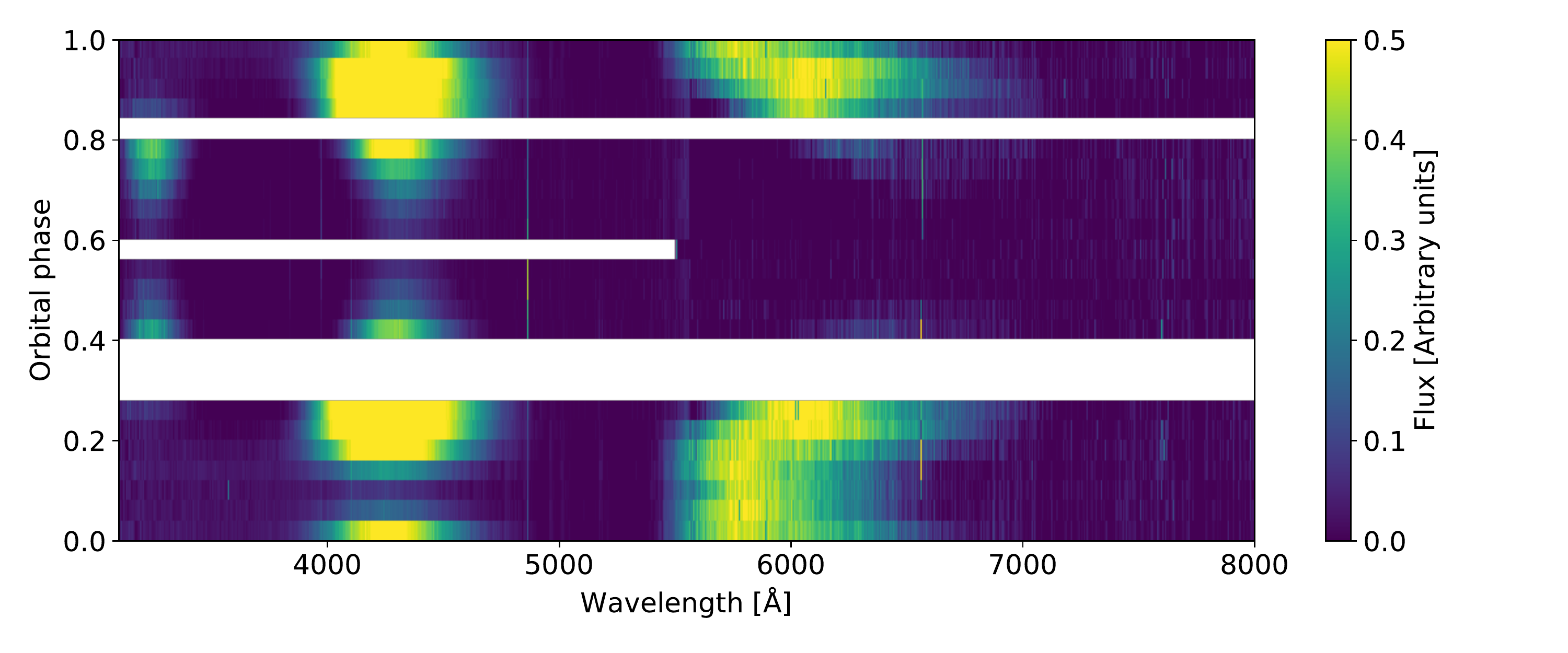}
    \caption{Phase-folded trailed spectra of SDSS\,J2229+1853 with the M dwarf component removed.}
  \label{fig:SDSS2229_trail}
  \end{center}
\end{figure*}

\subsection{White dwarf magnetic field strengths}

We estimate the magnetic field strengths of the white dwarfs in all the magnetic systems using the wavelengths of the cyclotron lines. The wavelength, $\lambda$, of the cyclotron harmonic number $n$ is related to the magnetic field $B$ via
\begin{equation}
\lambda_n = \left( \frac{10700}{n} \right) \left( \frac{10^8}{B} \right)\,$\AA$, \label{eqn:cyclo}
\end{equation}
where $B$ is measured in Gauss. This equation is valid for the plasma temperatures expected for white dwarf atmospheres \citep{Wickramasinghe00}. In general the strongest cyclotron lines are easy to see in the spectra, but field estimates usually require the identification of at least two cyclotron lines in order to uniquely determine the harmonic numbers. Since the M dwarfs typically dominate the spectrum, they can overwhelm weaker lines causing them to be overlooked. Therefore, we removed the M dwarf contributions from all the spectra, using the best fit template spectrum for the rotational broadening measurements. Figures~\ref{fig:SDSS0750_trail}-\ref{fig:SDSS2229_trail} show the resulting M dwarf subtracted spectra, plotted as trailed spectra. This process revealed weaker lines in many systems and allowed us to uniquely determine the field strengths via Equation~\ref{eqn:cyclo} in most systems. Unfortunately there is only one clear cyclotron line visible in the spectra of SDSS\,J0750+4943 and SDSS\,J1452+2045. For SDSS\,J1452+2045 the lack of any other visible cyclotron lines in the optical allows us to determine the harmonic number of the only visible line and hence the field strength, but for SDSS\,J0750+4943 there is some ambiguity in the harmonic number of the line, meaning that there are two possible fields strengths for this system. Some magnetic white dwarfs in our sample show poles with different field strengths, indicated in Table~\ref{tab:params}.

\section{Candidate magnetic systems} \label{sec:new_cands}

During a search for quasars in the SDSS spectroscopic database seven new candidate magnetic white dwarfs in detached binaries were serendipitously discovered. The SDSS spectra of these systems are plotted in Figure~\ref{fig:candidates}. All of these systems show broad emission features similar to cyclotron lines. The wavelengths of the cyclotron harmonics, estimates of the magnetic field strengths implied by these as well as {\it Gaia} early DR3 astrometric data are listed in Table~\ref{tab:candidates}. For the two systems where only a single cyclotron harmonic is detected we are unable to uniquely identify the harmonic number of the cyclotron lines, although the lack of any other cyclotron features means that these must be either the $n=1$ or $n=2$ harmonics. We also analysed the Zwicky Transient Facility (ZTF) Public Data Release 3 light curves of these systems \citep{Masci19}. Three systems show clear, periodic variability, likely related to the orbital period of the systems. However, we consider these periods preliminary. Follow up data are required in order to confirm that these are indeed magnetic white dwarfs in detached binaries and to establish the binary and stellar parameters, although the faintness of these systems makes this challenging. 

\begin{figure}
  \begin{center}
    \includegraphics[width=\columnwidth]{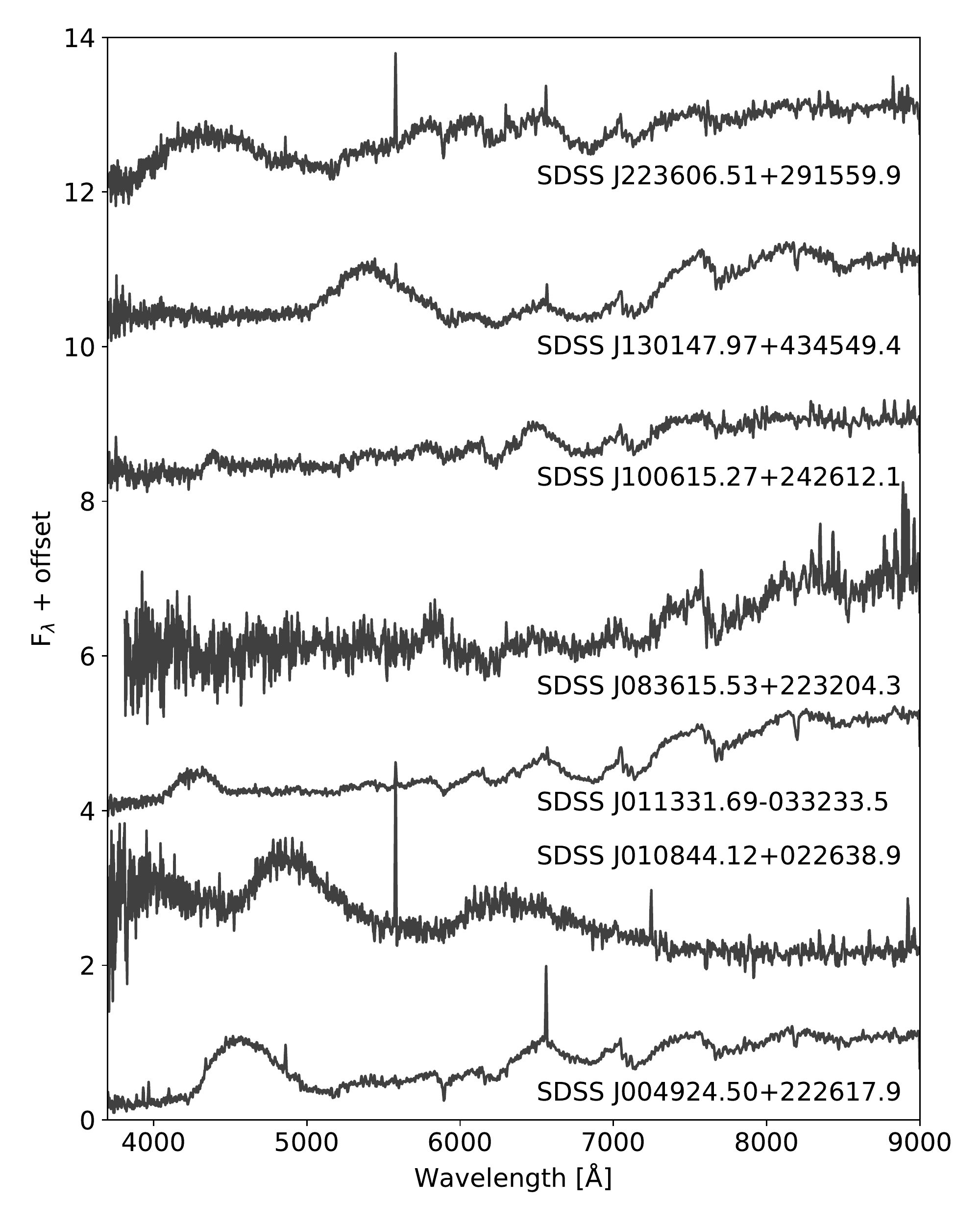}
    \caption{SDSS spectra of seven candidate close, detached binaries containing magnetic white dwarfs. The spectra are phase-average and have been smoothed by a 5-pixel boxcar. All of these systems appear to show strong cyclotron features. See Table~\ref{tab:candidates} for details of these systems.}
  \label{fig:candidates}
  \end{center}
\end{figure}

\begin{table*}
 \centering
  \caption{Candidate close, detached binaries containing magnetic white dwarfs from SDSS spectroscopy. The {\it Gaia} $G$ band magnitudes and parallaxes are taken from early DR3. The magnetic field strengths are purely based on the measured wavelengths of the cyclotron harmonics and were derived using Equation~\ref{eqn:cyclo}, as such they should be considered estimates. The photometric variability column lists any clear periodicity seen in the ZTF light curves of these sources.}
  \label{tab:candidates}
  \tabcolsep=0.1cm
  \begin{tabular}{@{}lccccc@{}}
  \hline
  Name & $G$   & $\pi$ & B    & Cyclotron harmonics & Photometric \\
       & (mag) & (mas) & (MG) & detected (\AA)      & variability \\
  \hline
  SDSS\,J004924.50+222617.9 & 18.52 & $1.53\pm0.25$ & 78 & 6840, 4560 ($n=2,3$) & 355 mins \\
  SDSS\,J010844.12+022638.9 & 20.60 & -             & 57 & 6450, 4850 ($n=3,4$) & 81 mins  \\
  SDSS\,J011331.69-033233.5 & 19.33 & $2.67\pm0.34$ & 84 & 6380, 4250 ($n=2,3$) & 263 mins \\
  SDSS\,J083615.53+223204.3 & 20.89 & -             & 184 or 92 & 5830 ($n=1$ or $n=2$) & No data  \\
  SDSS\,J100615.27+242612.1 & 20.02 & $0.42\pm0.53$ & 82 & 6530, 4350 ($n=2,3$) & No data  \\
  SDSS\,J130147.97+434549.4 & 19.96 & $2.19\pm0.43$ & 198 or 99 & 5400 ($n=1$ or $n=2$) & No data  \\
  SDSS\,J223606.51+291559.9 & 19.67 & $0.60\pm0.39$ & 83 & 6450, 4300 ($n=2,3$) & Non-variable\\
  \hline
\end{tabular}
\end{table*}

\section{Notes on individual systems} \label{sec:sys}

In this section we outline results for the systems studied in detail by us using X-shooter and/or IDS spectroscopy.

\subsection{Non-magnetic systems}

\subsubsection{SDSS\,J0225+0054}

With a period of 21.9 hours \cite{Nebot11}, this is one of the longest period systems observed in our sample. The system is well detached and shows no signs of magnetism in its spectrum, although our data only cover a limited phase range. However, given the low temperature of the white dwarf ($6600\pm300$\,K) it is perhaps unsurprising that it appears as a DC white dwarf in SDSS. The stellar parameters are relatively similar to the eclipsing system SDSS\,J1210+3347 \citep{Pyrzas12}, which hosts a DZ white dwarf showing a large number of metal absorption lines. The much longer period of SDSS\,J0225+0054 likely means that the accretion rate onto the white dwarf is too low to generate detectable metal lines. The long period of this system also means that the rotational broadening of the M star is small ($5.5\pm2.3$\,\kms) leading to a substantial uncertainty on the white dwarf mass. 

\subsubsection{SDSS\,J0848+2320}

The SDSS spectra of this system showed substantial radial velocity variations indicating that it was a close binary. Our X-shooter data revealed an orbital period of 8.9 hours. No Balmer lines are seen from the white dwarf, but clear Ca\,{\sc ii} absorption is seen at 3934{\AA} moving in anti-phase with the M dwarf features, indicating this line originates from the white dwarf. This system therefore made a excellent target to check the reliability of using the rotational broadening to constrain the mass ratio (hence white dwarf mass), since we have an independent measurement of the mass ratio directly from the radial velocity data. The mass ratio derived from the rotational broadening ($q=0.94\pm0.06$) is in good agreement with the directly measured value from the radial velocities ($q=0.97\pm0.03$), which lends confidence to the parameters derived via this technique in the other systems.

The spectra show no clear evidence of magnetism from the white dwarf. The sharpness of the Ca\,{\sc ii} absorption line also argues against a strongly magnetic white dwarf in this system. 

\subsubsection{SDSS\,J1140+1542}

The SDSS spectra of this system showed some radial velocity variations indicating that it was a close binary. Our X-shooter data confirmed that this is indeed the case and, when combined with previous velocity measurements, yielded a period of 3.1 days, by far the longest period system in our sample. As such, the system is well detached and shows no evidence of magnetism from the white dwarf, although our phase coverage is quite sparse. The very small rotational broadening in this system means that it cannot place useful constraints on the white dwarf mass. However, the minimum white dwarf mass implied by the orbital period, radial velocity semi-amplitude and mass of the M dwarf is 1.22\,{\MSUN}, making this one of the most massive white dwarfs in a detached PCEB.

\subsubsection{SDSS\,J1316-0037}

This 9.7 hour binary \citep{Nebot11} shows no obvious features from the white dwarf in its X-shooter spectrum. It is well detached, with no clear signs of magnetism from the white dwarf, although our phase coverage is low. We note that there is a slight discrepancy between the M dwarf mass that we report and the value in \citet{Nebot11}. The mass reported in \citet{Nebot11} is based on a mass-spectral type relationship \citep{Rebassa07}, which does not yield precise masses and should only be considered an estimate. Our M dwarf masses are based on the $K_s$ band luminosities of these stars, which yields more precise and accurate masses when compared to independent measurements \citep{Mann19,Parsons18}.

\subsubsection{SDSS\,J2208+0037}

With a period of only 2.5 hours \citep{Zorotovic16} this system sits right in the middle of the CV period gap. This system is much closer to Roche-lobe filling than the other non-magnetic systems. Interestingly, this system has many traits in common with the magnetic systems, such as a short period and high Roche-lobe filling factor, as well as a cool white dwarf ($6100\pm400$\,K). The only obvious difference is the low white dwarf mass $0.46\pm0.24$\,{\MSUN}, which is substantially smaller than the white dwarf masses measured in the magnetic systems. Despite the similarities with the magnetic white dwarf systems, there is no clear evidence of magnetism from the white dwarf in the X-shooter data.

\subsection{Magnetic systems}

\subsubsection{SDSS\,J0750+4943}

This system was initially identified as a white dwarf plus main-sequence binary from its SDSS colours \citep{Rebassa13} and found to be a close binary with a period of 4.2 hours from its CRTS light curve \citep{Parsons15}. The shape of the CRTS light curve was reminiscent of those with strongly Roche-distorted M dwarfs, but was highly asymmetric, distinguishing it from most other systems. Our spectroscopic observations revealed that this system contains a magnetic white dwarf and the strong cyclotron line at $\sim$5450\,{\AA} (which is only visible for half of the period) was the cause of the asymmetric light curve. This demonstrates that these detached magnetic systems can be potentially identified through their unusual light curves. Inspection of the spectra taken when the cyclotron emission is absent show no clear features from the white dwarf.

We were able to measure a radial velocity semi-amplitude of the M dwarf from our INT/IDS spectra, but not a rotational broadening. However, SDSS\,J0750+4943 is the only system in our sample with both a precise Gaia parallax and high quality ultraviolet fluxes from the Galaxy Evolution Explorer ({\it GALEX}) mission \citep{Martin05}. As such, we constrained both the surface gravity and temperature of the white dwarf by fitting its ultraviolet fluxes. We used non-magnetic white dwarf models \citep{Koester10} and assumed a standard mass-radius relationship for carbon-oxygen core white dwarfs \citep{Fontaine01} in order to scale the model spectrum. We fitted the data using the Markov Chain Monte Carlo method \citep{Press07} implemented using the python package {\sc emcee} \citep{Foreman13}. We included a prior on the parallax based on the Gaia measurement and a prior on the reddening of $E(B-V)=0.023\pm0.019$ mags, based on Gaia-2MASS dust maps \citep{Lallement19}. The result of this fit is shown in Figure~\ref{fig:SDSS0750_SED}.

We note that it is unclear how strong any cyclotron emission is in the ultraviolet. Because only a single cyclotron line is visible in the IDS spectrum we can only constrain the white dwarf field strength to be either 196\,MG (the 5450\,{\AA} line is the fundamental) or 98\,MG (the 5450\,{\AA} line is the $n=2$ harmonic). We can exclude the possibility of the cyclotron line being a higher harmonic as this would result in additional harmonics appearing in the IDS spectrum, which are not seen. If the field strength is the higher value then several cyclotron harmonics are likely to contribute to the ultraviolet flux. On the other hand this is less likely to be an issue if the field strength is the lower value. This source was observed by {\it GALEX} in the NUV band for roughly 30 minutes. Inspection of the light curve of the system (produced using the python package {\sc gphoton}, \citealt{Million16}) over the 30 minute timespan shows not obvious variation, which might be expected if there was a substantial cyclotron component. However, this only covers a small fraction of the orbit so we cannot rule out contamination by cyclotron emission. Therefore, given this, and the fact that we modelled the white dwarf using non-magnetic white dwarf models, the parameters of this system should be interpreted with great caution.

\begin{figure}
  \begin{center}
    \includegraphics[width=\columnwidth]{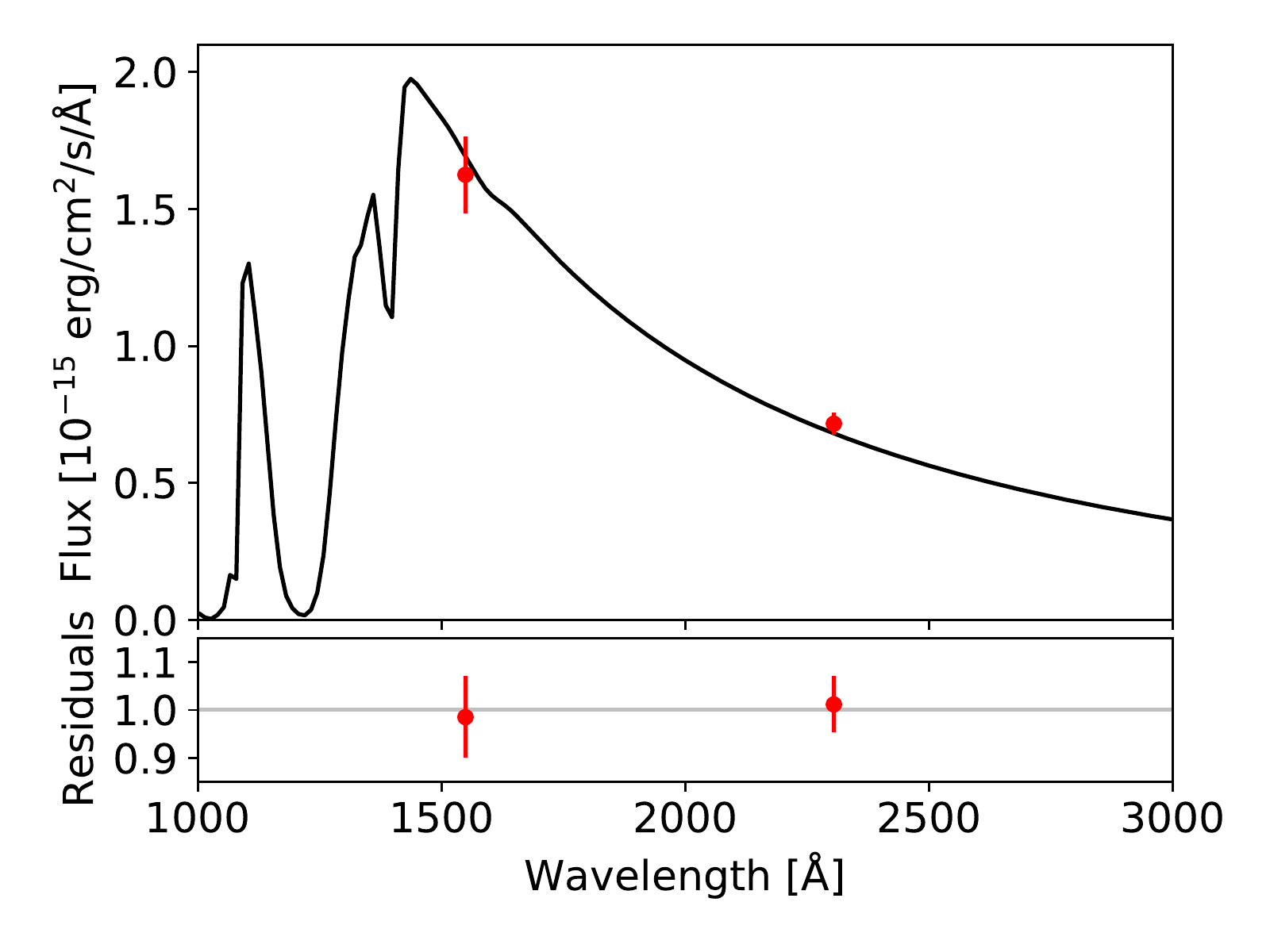}
    \includegraphics[width=\columnwidth]{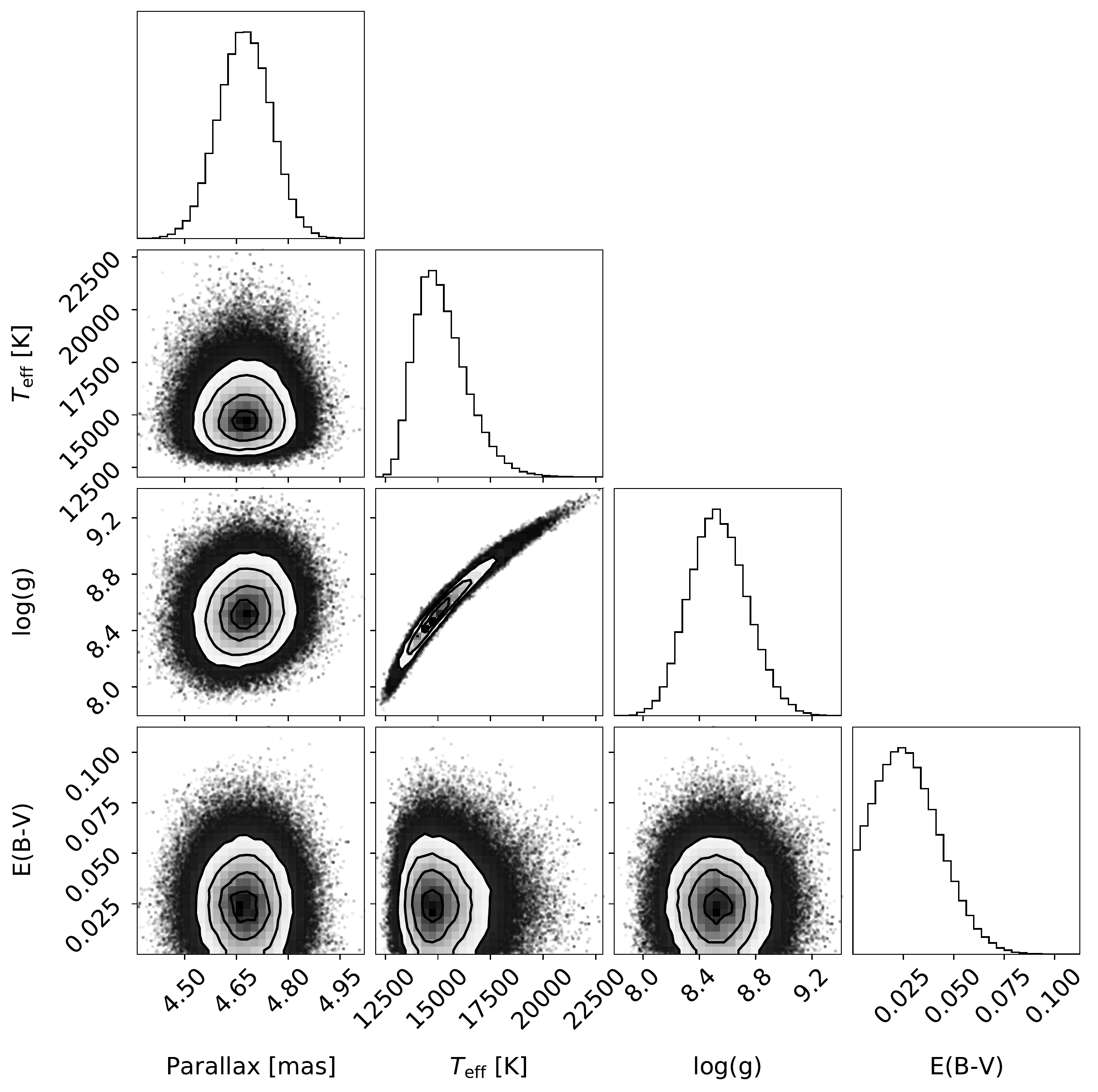}
    \caption{{\it Top:} Fit to the {\it GALEX} FUV and NUV measurements for the magnetic system SDSS\,J0750+4943 using non-magnetic DA white dwarf models. {\it Bottom:} Posterior probability distributions for model parameters obtained from the fit. Priors we placed on both the parallax and reddening.}
  \label{fig:SDSS0750_SED}
  \end{center}
\end{figure}

\subsubsection{SDSS\,J0853+0720}

This system was previously classified as a DC+dM binary with a period of 3.6 hours \citep{Nebot11}. The SDSS spectrum showed no clear evidence that the white dwarf was magnetic. However, our X-shooter data revealed two cyclotron lines. A strong line is seen at 4250\,{\AA} and a weaker line is evident at 6300\,{\AA} once the M dwarf component was subtracted (see Figure~\ref{fig:SDSS0853_trail}), which are the the $n=3$ and $n=2$ harmonics of an 84\,MG field. Both of these cyclotron lines become faint after phase 0.5 and so it is likely that the SDSS spectrum was obtained around this phase and thus missed the cyclotron lines. No clear white dwarf features are visible in the X-shooter spectra at the phase when the cyclotron emission is weakest. This result confirms our suspicion that magnetic white dwarfs could be hiding as DC white dwarfs within the population of white dwarf plus main-sequence star systems discovered in SDSS. This is further reinforced by the temperature estimate of $9000\pm300$\,K for this white dwarf. At this temperature the white dwarf should still show strong Balmer absorption lines, even against the M dwarf. Zeeman splitting has resulted in these lines becoming too weak to be detected against the M dwarf.

\subsubsection{HS\,0922+1333}

Originally discovered in the Hamburg Quasar Survey \citep{Reimers00}, HS\,0922+1333 was the second detached magnetic white dwarf system discovered. It shows evidence of two poles with different field strengths of 66\,MG (visible in Figure~\ref{fig:HS0922_trail} around phase 0.9 where the $n=2$ harmonic at 8100\,{\AA}, $n=3$ harmonic at 5400\,{\AA} and $n=4$ harmonic at 4050\,{\AA} are all visible) and 81\,MG (visible around phase 0.5 where the $n=3$ harmonic at 4400\,{\AA} is seen along with a weak $n=2$ component at 6600\,{\AA}). The orbital period was refined by \citet{Tovmassian07}, who also found tentative evidence of Roche-lobe overflow via a stream of material leaving the L$_1$ point. We see no such feature in our X-shooter data, although our data cover less than half the orbit, so it is conceivable that we missed this feature.  

Our new data improve the radial velocity semi-amplitude measurement for the M dwarf and, for the first time, yield a measure of its rotational broadening. Using these measurements we find a white dwarf mass of $0.71\pm0.07$\,{\MSUN}, somewhat higher than the average white dwarf mass in detached white dwarf plus main-sequence binaries, but fairly typical for white dwarfs in CVs \citep{Zorotovic11}.

\subsubsection{IL\,Leo} \label{sec:IL_Leo}

\begin{figure}
  \begin{center}
    \includegraphics[width=\columnwidth]{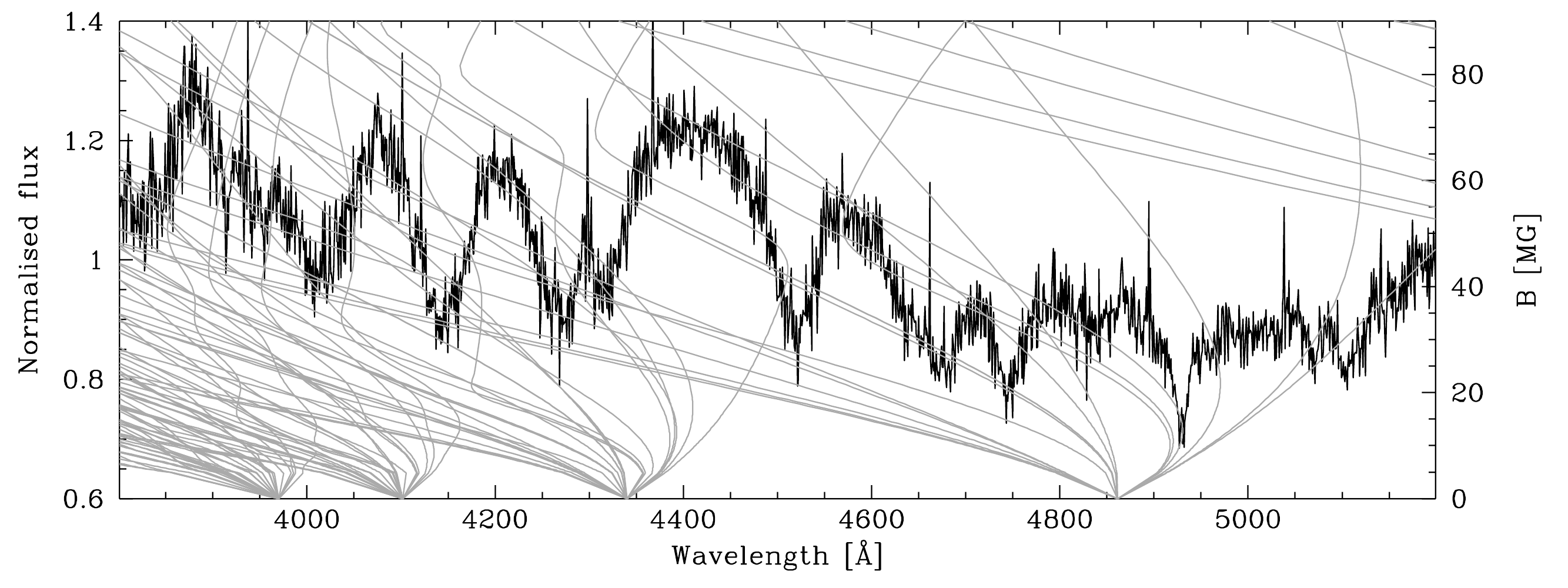}
    \caption{Average spectrum of IL\,Leo (black line). Multiple Zeeman split hydrogen Balmer absorption lines from the white dwarf are seen. The wavelengths of the Zeeman split components \citep{Friedrich96} are shown in grey as a function of the magnetic field strength. The field strength of the white dwarf measured from the cyclotron emission is 42\,MG. The $n=5$ (5160\,\AA) and $n=6$ (4300\,\AA) cyclotron harmonics are visible in this plot, although they are heavily distorted by the absorption lines.}
  \label{fig:ILLeo_av}
  \end{center}
\end{figure}

Also known as SDSS\,J1031+2028, this system was originally discovered from its unusual SDSS spectrum after being targeted as a quasar candidate \citep{Schmidt07}. IL\,Leo is unique among our sample as the only system where the main-sequence component does not dominate the optical flux. In fact, the faint H$\alpha$ emission line shown in Figure~\ref{fig:ILLeo_RV} is the only features visible in the X-shooter data from the main-sequence star. Furthermore, IL\,Leo is the only magnetic system in our sample that clearly shows Zeeman split Balmer absorption lines (see Figure~\ref{fig:ILLeo_av}). The X-shooter data show strong cyclotron emission from a 42\,MG magnetic field. The $n=3$ harmonic at 8600\,{\AA}, the $n=4$ harmonic at 6450\,{\AA}, $n=5$ harmonic at 5160\,{\AA} and $n=6$ harmonic at 4300\,{\AA} are all visible (best seen in Figure~\ref{fig:ILLeo_trail}). The wavelengths of the Zeeman split Balmer absorption lines appear consistent with this field strength, but the high field strength implies that many Zeeman components will be smeared out by the variation of field strength across the white dwarf surface (factor of $\simeq$2 for a dipole configuration). It is therefore difficult to measure the field strength directly from the absorption lines, thus we adopt the field strength as measured from the cyclotron emission.

The lack of a direct radial velocity semi-amplitude for the main-sequence star (the H$\alpha$ emission only places a lower limit on this) and no rotational broadening measurement makes it difficult to place any meaningful constraints on the stellar parameters. However, the short period of the system, just 82 minutes, does offer some limits on the system parameters. 

Assuming that the main-sequence star follows the same mass-radius relationship from Section~\ref{sec:md_mr} then it must be less massive than 0.09\,{\MSUN} in order to fit within its Roche-lobe at this orbital period, placing the star close to the brown dwarf regime. Moreover, the Roche-lobe filling factor must be $>$0.95. Smaller filling factors than this result in star that is substantially undersized for its mass. This is true for any mass above $\sim$0.07\,{\MSUN}, below this mass the radius is sensitive to the age of the system and in principle a very old ($>$5\,Gyr) brown dwarf could have a smaller filling factor than this. However, we consider the possibility of such a low mass companion unlikely, since the strength of the cyclotron lines imply that the white dwarf is capturing a substantial amount of wind material. Brown dwarfs are not expected to produce strong winds and so it is unlikely that a brown dwarf, well within its Roche-lobe, could generate enough wind material to produce the strong cyclotron features seen from the white dwarf. 

The H$\alpha$ emission line from the companion seen in the X-shooter data of IL\,Leo (Figure~\ref{fig:ILLeo_RV}) may be driven by irradiation from some accretion source in the system, since the temperature of the white dwarf is too low to generate such a strong emission line (see for example the white dwarf plus brown dwarf binary WD\,1032+011, which has a slightly longer period but contains a similar temperature white dwarf to IL\,Leo, but shows no evidence of any emission lines from the brown dwarf \citealt{Casewell20}). This may be evidence of Roche-lobe overflow in IL\,Leo, although there is no clear emission from an accretion stream in the X-shooter data.

Finally, for a given set of stellar masses and a Roche-lobe filling factor, the radial velocity of the H$\alpha$ emission line can be corrected to the centre-of-mass of the main-sequence star following the technique outlined in \citet{Parsons12_ushort}. This places a lower limit on the mass of the white dwarf, below which the measured radial velocity would imply an unfeasible inclination regardless of the companion mass (more specifically, the system should be eclipsing, which IL\,Leo is not). Unfortunately the limit on the white dwarf mass is only $>$0.48\,{\MSUN}, which is not very informative.

\subsubsection{SDSS\,J1452+2045}

The period of this system was measured by \citet{Zorotovic16} to be 2.6 hours, placing it right in the middle of the period gap. The SDSS spectrum of this system shows no evidence of magnetism and the system was classified as a DC+dM binary. However, our X-shooter data revealed a strong double-peaked cyclotron line at around 3600\,{\AA}, which is outside of the SDSS spectral range, hence the reason that the magnetic nature of the white dwarf was previously missed. Remarkably, no other cyclotron lines are visible implying that the line at 3600\,{\AA} is the fundamental from a 300\,MG field (if this line was a higher harmonic then additional cyclotron lines should be visible in the X-shooter data). This is comfortably the highest field strength among our sample and is higher than any known white dwarf in a CV \citep{Ferrario15}, although there are a number of single magnetic white dwarfs with larger field strengths than this. No clear white dwarf features are visible in the spectra at any orbital phase.

This system, along with the 8\,MG eclipsing system SDSS\,J0303+0054, demonstrates a key limitation of identifying magnetic systems from cyclotron emission at optical wavelengths. Only a limited range of field strengths will create strong cyclotron emission at optical wavelengths. Very high fields (e.g. SDSS\,J1452+2045) will show cyclotron emission in the ultraviolet, while very low field strengths (e.g. SDSS\,J0303+0054) will only show cyclotron emission at infrared wavelengths. 

With a mass of $0.83\pm0.08$\,{\MSUN} the white dwarf in SDSS\,J1452+2045 is more massive than those typically found in detached binaries with main-sequence stars \citep{Zorotovic11}. However, assuming that the main-sequence star follows the same mass-radius relationship from Section~\ref{sec:md_mr} then the system is quite well detached (filling factor of 0.84).

\subsubsection{SDSS\,J2229+1853}

Like SDSS\,J0750+4943 this system was found to be a close binary with an unusual asymmetric light curve \citep{Parsons15}. Our follow-up data confirm the magnetic nature of the white dwarf via the identification of very strong cyclotron lines. The $n=2$ harmonic at 6400\,{\AA}, $n=3$ harmonic at 4250\,{\AA} and $n=4$ harmonic at 3200\,{\AA} from an 84\,MG field are visible in Figure~\ref{fig:SDSS2229_trail}), although the field is clearly complex as the $n=2$ harmonic weakens as $n=4$ harmonic strengthens. Higher phase resolution or spectropolarimetric observations are required to better understand the structure of the magnetic field of the white dwarf in this system. With a period of 4.5 hours SDSS\,J2229+1853 has the longest period of all known detached magnetic white dwarf binaries and appears to be extremely close to Roche-lobe filling. No clear white dwarf features are visible in the X-shooter spectra at any orbital phase.

\section{Discussion}

\subsection{The search for low field detached magnetic white dwarf binaries}

Our target selection was driven by the discovery of the low field magnetic white dwarf in SDSS\,J0303+0054 \citep{Parsons13_mag}. The spectrum of the white dwarf in this binary lacks any hydrogen Balmer absorption lines, despite its temperature being high enough to produce lines that should be visible in the SDSS spectrum for example. Detailed analysis showed that the lack of absorption lines is the result of a combination of Zeeman splitting (which weakens the lines) and an additional (Zeeman split) emission component originating from the white dwarf, which fills the absorption lines. Given that emission lines are often seen from cool white dwarfs accreting from the wind of companion stars \citep[e.g.][]{Tappert11,Parsons12_ucool,parsons17,Longstaff19} we hypothesised that there could be other magnetic white dwarfs hiding among the sample of white dwarf plus main-sequence star binaries classified as DC+dM systems. Importantly, these systems would have weaker magnetic fields than those seen in previously discovered detached magnetic white dwarf binaries since the SDSS spectra of DC+dM binaries show no evidence of cyclotron emission. Like SDSS\,J0303+0054, this emission would instead occur at infrared wavelengths.

Our high-quality X-shooter spectra of DC+dM binaries did reveal two previously unknown magnetic systems (SDSS\,J0853+0720 and SDSS\,J1452+2045), but the magnetic nature of the white dwarf in both of these binaries was revealed via cyclotron emission (which is not seen in the SDSS spectra of these sources), rather than Zeeman split emission from the white dwarf. In fact, SDSS\,J0303+0054 remains the only detached magnetic white dwarf binary to show Zeeman split emission. None of the other DC+dM binaries show any emission lines from the white dwarf (either Zeeman split or not), leading us to conclude that they are unlikely to possess even weak magnetic fields. The Balmer absorption lines from these white dwarfs are weak due to their low temperatures and are therefore undetectable next to their brighter main-sequence companions.

It is likely that the higher magnetic field strengths of white dwarfs in detected magnetic systems compared to SDSS\,J0303+0054 (10s of MG, compared to 8\,MG for SDSS\,J0303+0054) means that any hydrogen emission from the white dwarfs in these binaries is so strongly Zeeman split that it is undetectable. Hence the reason that the magnetic white dwarf in SDSS\,J0303+0054 is the only one to currently show Zeeman split emission lines. The low magnetic field strength of the white dwarf in SDSS\,J0303+0054 remains an outlier.

\subsection{General properties of detached magnetic white dwarf binaries}

The detached magnetic white dwarf systems share a number of common properties. In general they contain cool white dwarfs, with temperatures lower than the majority of detached non-magnetic white dwarfs and cooler than the white dwarfs in polars \citep{Schwope09}. Additionally, these detached magnetic white dwarf systems are all very close to Roche-lobe filling (see Figure~\ref{fig:age}). However, given that the magnetic nature of the white dwarfs in virtually all of these systems was uncovered via the identification of cyclotron emission, these may be selection effects. Wind accretion will be largest in those systems closest to Roche-lobe overflow, leading to stronger cyclotron lines. Moreover, the cyclotron lines are easier to see when the white dwarfs are cooler, since the cyclotron emission dominates over the photosphere. On the other hand, Zeeman splitting of the hydrogen Balmer lines would be easier to see in hotter white dwarfs and would also not depend upon having any wind accretion from the main-sequence star. Therefore, the lack of any magnetic white dwarfs hotter than $\sim$10000\,K among the thousands of white dwarf plus main-sequence star systems spectroscopically observed by SDSS \citep{Rebassa16} strongly implies that these systems are extremely rare or perhaps do not exist at all. Interestingly, of the 10 systems originally listed in \citet{Rebassa16} as DC+dM binaries that have been observed in detail with X-shooter, 4 are magnetic, implying that magnetic systems appear to be much more common among cooler, hence older, white dwarf systems. This is similar to the suggestion of an increase in the incidence of magnetism among cooler single white dwarfs \citep{Liebert03,Kawka14,Hollands15}

\begin{figure}
  \begin{center}
    \includegraphics[width=\columnwidth]{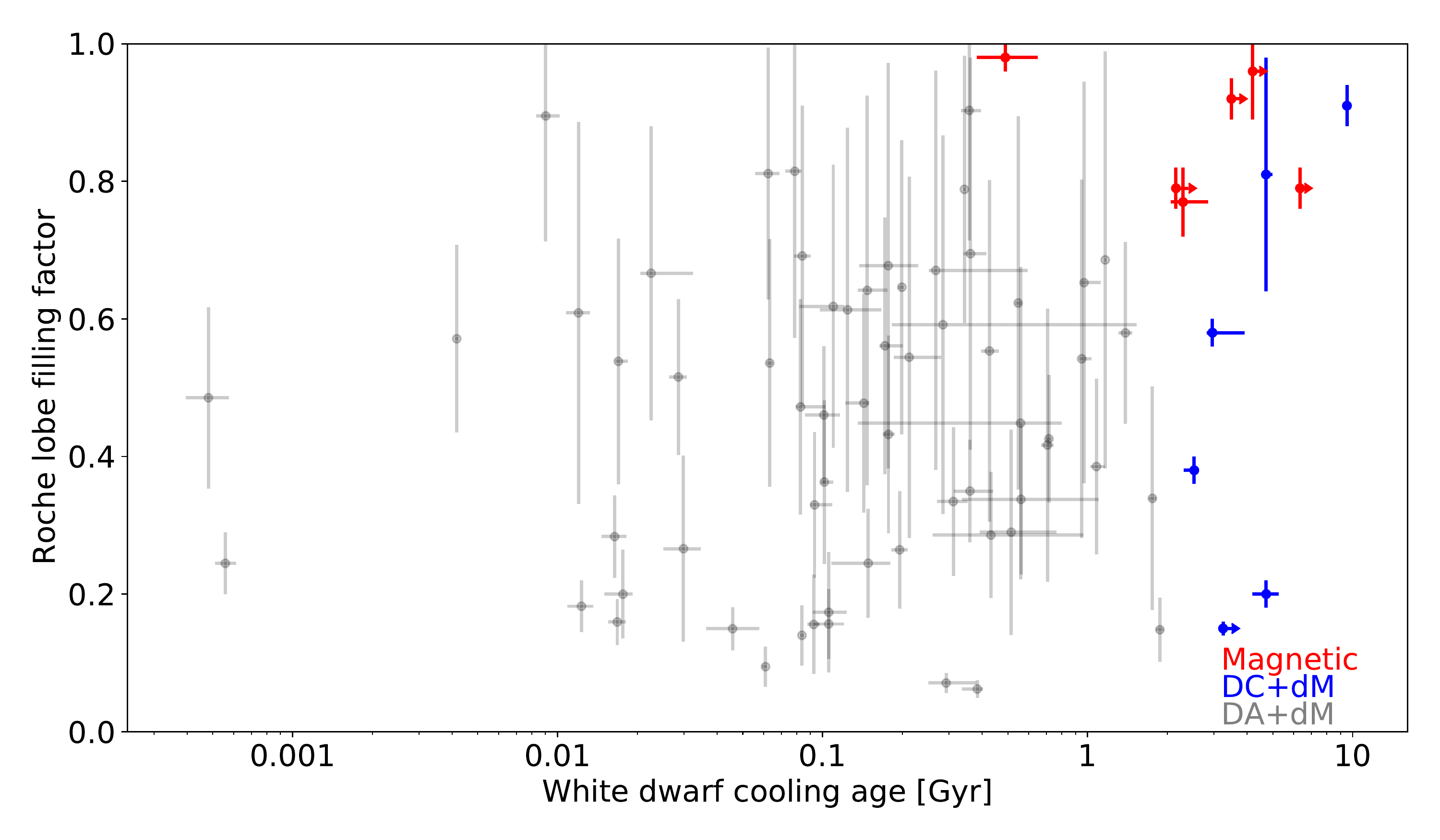}
    \caption{Roche-lobe filling factors for detached white dwarf plus main-sequence binaries as a function of the white dwarf cooling age. Arrows denote lower limits. DA+dM systems taken from \citet{Rebassa16}. Cooling ages were determined using the cooling sequence from \citet{Fontaine01}. On average magnetic systems have larger filling factors and longer cooling ages than non-magnetic systems.}
  \label{fig:age}
  \end{center}
\end{figure}

\begin{figure}
  \begin{center}
    \includegraphics[width=\columnwidth]{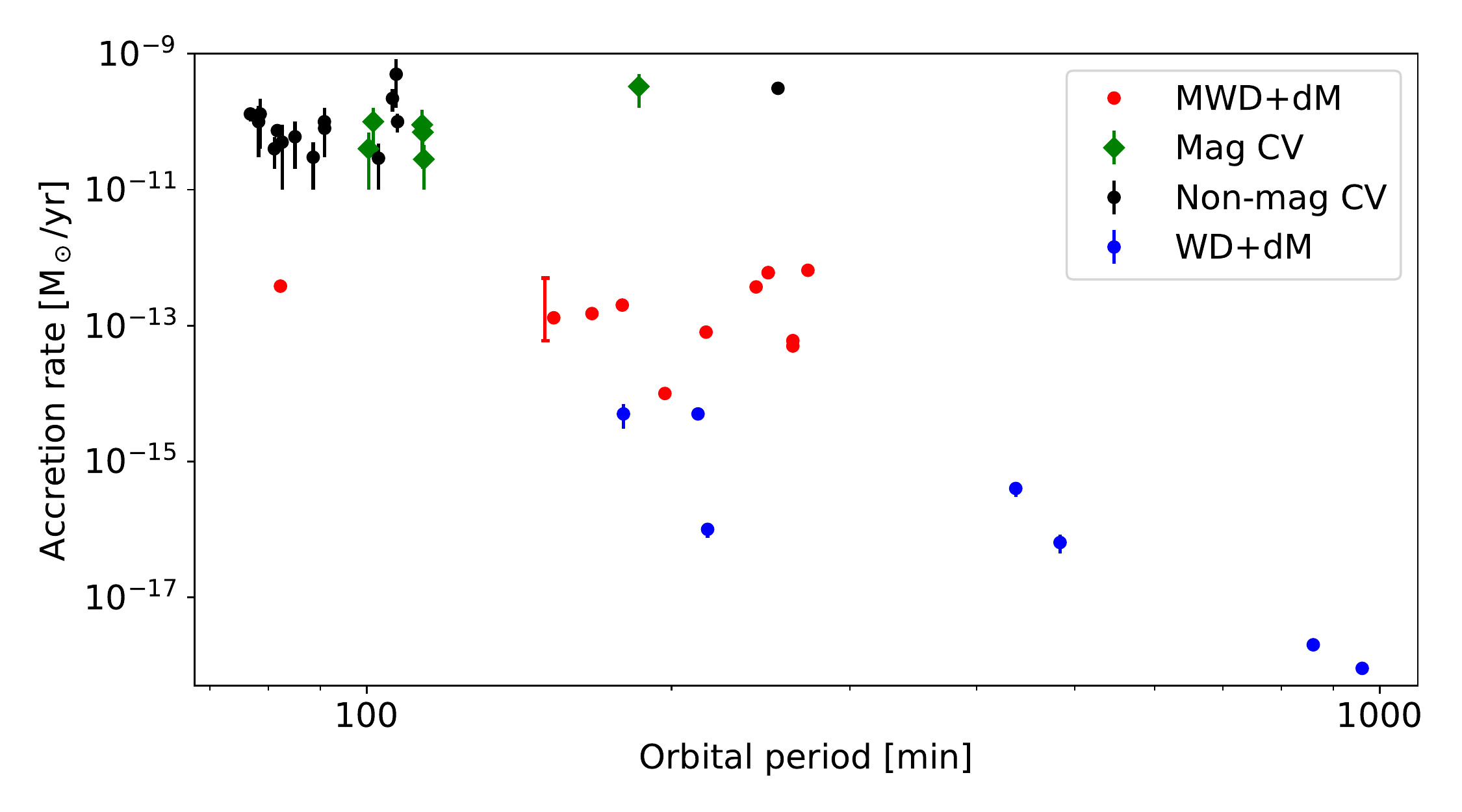}
    \caption{The measured accretion rates for a sample of nearby cataclysmic variables (non-magnetic in black, magnetic in green), detached magnetic white dwarf binaries (red) and detached non-magnetic white dwarf binaries (blue).}
  \label{fig:acc_rate}
  \end{center}
\end{figure}

\subsection{Accretion rates}

We estimated the accretion rate of material onto the white dwarfs in the magnetic systems following the approach of \citet{Schwope09} whereby the cyclotron luminosity is equated to the accretion luminosity. We use the Gaia distance estimates and measure the peak integrated flux of the cyclotron lines in our M dwarf subtracted spectra (using the spectra taken when the cyclotron lines are at their strongest). We removed any contribution from the white dwarf's photosphere by subtracting off a model white dwarf spectrum (non-magnetic, hydrogen atmosphere, \citealt{Koester10}) with a temperature set to our upper limits and scaled to the distance of the system. In most cases subtracting off this contribution has a negligible effect on the accretion rate estimates. Since our data cover such a wide wavelength range (with the exception of SDSS\,J0750+4943) it is likely that we have included most of the cyclotron flux. However, it is possible that there is significant flux in cyclotron lines outside of our wavelength range in some cases. Moreover, cyclotron emission is generally anisotropic and beamed adding further uncertainty to our accretion rate measurements. In addition, a fraction of the accretion luminosity may be emitted at X-ray wavelengths \citep[e.g.][]{Vogel07}. Hence the accretion rates should be interpreted as only approximate.

The estimated accretion rates are given in Table~\ref{tab:acc_rates}. Figure~\ref{fig:acc_rate} shows these accretion rates relative to those measured for CVs within 150\,pc \citep{Pala20} as well as detached binaries with non-magnetic white dwarfs \citep{Debes06,Tappert11,Pyrzas12,Parsons12,Drake14}. The accretion rates of these detached magnetic white dwarf binaries are clearly below those seen in CVs, but are substantially higher than the accretion rates seen in pre-cataclysmic binaries with non-magnetic white dwarfs.

The higher accretion rates seen in detached magnetic white dwarf binaries compared to non-magnetic white dwarf binaries with similar orbital periods implies that the magnetic white dwarfs are able to capture a significantly larger fraction of the wind material from their companion. This supports the idea that angular momentum loss via magnetic braking may be reduced in binaries containing a strongly magnetic white dwarf because the wind material from the companion is trapped within the magnetosphere of the white dwarf and hence is unable to carry away as much angular momentum \citep{Li94,Webbink02}. Population synthesis calculations have recently shown that reduced magnetic braking is required to reproduce the observed orbital period distribution, space density and mass transfer rates of magnetic CVs \citep{Belloni20_EMB}.

\begin{table}
 \centering
  \caption{Estimated accretion rates in detached magnetic white dwarf binaries. It is likely that our data do not capture the full cyclotron luminosity and therefore these accretion rates are only approximate.}
  \label{tab:acc_rates}
  \tabcolsep=0.11cm
  \begin{tabular}{@{}lccc@{}}
  \hline
  Name & $\dot{M}$ &  Reference \\
       & ($10^{-13}${\MSUN}/yr) & \\
  \hline
  SDSS\,J0750+4943 & 6.0 & This paper \\
  SDSS\,J0837+3830 & 2.0 & \citet{Schmidt05_2} \\
  SDSS\,J0853+0720 & 0.8 & This paper \\
  HS\,0922+1333    & 3.7 & This paper \\
  WX\,LMi          & 1.5 & \citet{Vogel07} \\
  IL\,Leo          & 3.8 & This paper \\
  SDSS\,J1059+2727 & $0.6-5.0$ & \citet{Schmidt07} \\
  SDSS\,J1206+5100 & 0.1 & \citet{Schwope09} \\
  SDSS\,J1452+2045 & 1.3 & This paper \\
  MQ\,Dra          & 0.6 & \citet{Schmidt05_2} \\
  SDSS\,J2048+0050 & 0.5 & \citet{Schmidt05_2} \\
  SDSS\,J2229+1853 & 6.5 & This paper \\
  \hline
\end{tabular}
\end{table}

\subsection{Evolutionary status}

The evolutionary status of detached binaries containing magnetic white dwarfs has been addressed by many authors \citep[e.g.][]{Schwope09,Breedt12}, but a general consensus has yet to be reached. The largest question remains whether these systems are CVs that are in extreme low states or have detached all together (low accretion rate polars, LARPs), or if they are pre-CVs, where the main-sequence star is yet to fill its Roche-lobe (pre-polars, PREPs). Several systems appear to show signs of episodic Roche-lobe overflow. For example, \citet{Tovmassian07} saw evidence of a high velocity H$\alpha$ component in HS\,0922+1333, which they interpret as an accretion stream, although no such feature is evident in our X-shooter data. A similar feature was found in SDSS\,J2048+0050 by \citet{Kafka10}, which was not seen in previous data \citep{Schmidt05_2}. Inspection of the long term CRTS light curves of these systems also reveals a clear brightening event in SDSS\,J0837+3830 (see Figure~\ref{fig:SDSS0837_lc}), where it is likely that the accretion rate increased dramatically for around a year, likely as a result of Roche-lobe overflow, before returning to its original brightness. It therefore appears more likely that these systems are CVs that have temporarily detached. This is also supported by our white dwarf mass estimates. For the 6 detached magnetic systems with reliable mass estimates we obtain an average white dwarf mass of $\langle M_\mathrm{WD} \rangle = 0.82 \pm 0.07$\,{\MSUN}, which is  higher than the average mass of white dwarfs in pre-CVs ($0.67\pm0.21$\,{\MSUN}, \citealt{Zorotovic11}), but is consistent with both non-magnetic CV white dwarf masses ($0.83\pm0.23$\,{\MSUN}, \citealt{Zorotovic11,Schreiber16}) and magnetic CV white dwarf masses ($0.77\pm0.02$\,{\MSUN}, \citealt{Shaw20}, $0.84 \pm 0.17$\,{\MSUN} \citealt{Martino20}). Although it is worth noting that in general isolated magnetic white dwarfs appear to have higher masses than their non-magnetic counterparts \citep{Ferrario15,McCleery20}

\begin{figure}
  \begin{center}
    \includegraphics[width=\columnwidth]{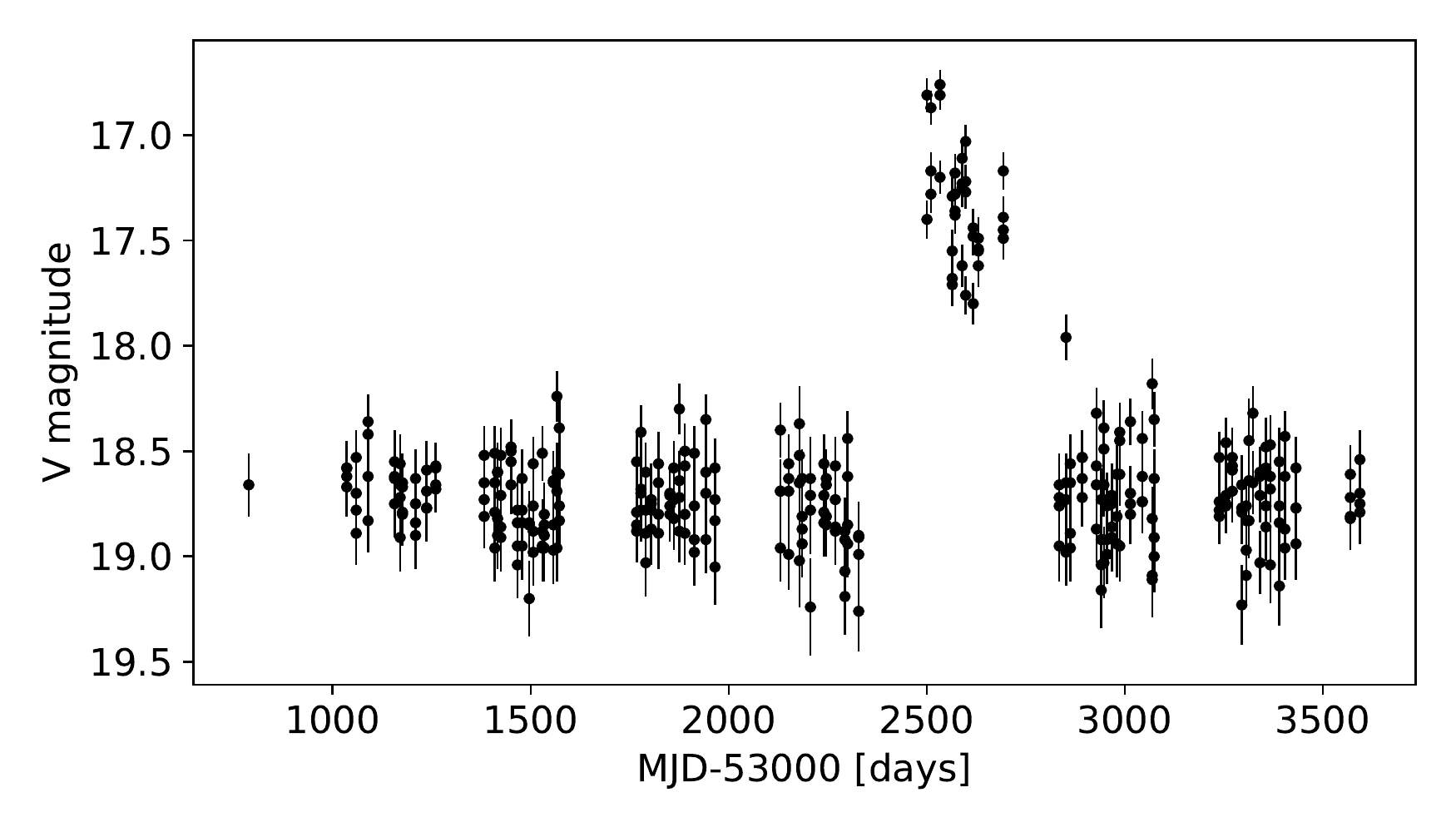}
    \caption{CRTS light curve of the magnetic system SDSS\,J0837+3830 showing a period of increased brightness likely as a result of accretion via Roche-lobe overflow.}
  \label{fig:SDSS0837_lc}
  \end{center}
\end{figure}

On the other hand, our measured Roche-lobe filling factors are as low as 0.84, which would place the main-sequence star quite far from Roche-lobe filling. Indeed, it will take almost 1\,Gyr for SDSS\,J0303+0054 to come into contact \citep{Parsons13_mag}, which would argue more towards a pre-CV classification for this system. However, the donor stars in CVs are pushed out of thermal equilibrium and thus have radii larger than isolated stars of the same mass. Therefore, if the system becomes detached long enough for the donor to relax back to thermal equilibrium then it can substantially underfill the Roche-lobe (this is thought to be the cause of the orbital period gap for example). Figure~\ref{fig:max_detach} shows what the expected Roche-lobe filling factors that CVs would drop to if the donor star was allowed to relax down to an equilibrium radius (black line), using the standard evolutionary track from \citet{Knigge11} and an equilibrium radius given by the mass-radius relation from Section~\ref{sec:md_mr}. Hence any system on or above this line could, in principle, have been Roche-lobe filling in the past and the donor has now relaxed. All of the magnetic systems with reliable Roche-lobe filling factors are on or above this line, hence there are no systems were we can rule out a previous phase of mass transfer via Roche-lobe overflow. Also note that the radii of the main-sequence stars in all these systems were determined from the semi-empirical mass-radius relationship and hence we have assumed that the main-sequence stars in these systems are in thermal equilibrium. If this is not the case then our Roche-lobe filling factors will be underestimated, hence the points in Figure~\ref{fig:max_detach} are likely lower limits, rather than exact values (the exception being the eclipsing system SDSS\,J0303+0054, which has a direct radius measurement). 

Although we cannot rule out the possibility that these systems are in fact pre-CVs, all the evidence supports the interpretation that these are CVs that are in extreme low states or completely detached. This would imply that we are yet to identify any magnetic white dwarfs in pre-CVs, hence the origin of magnetic CVs remains unclear.

It is worth noting that the population of detached magnetic white dwarfs is still very biased. The vast majority of these systems were identified from cyclotron lines in their optical spectra, which are stronger in systems closer to Roche-lobe filling for example. Cyclotron emission is also easier to detect when the photosphere of the white dwarf is cool and hence faint, biasing us towards systems with cooler white dwarfs. On the other hand, Zeeman splitting of the hydrogen Balmer lines is easier to identify in hotter white dwarfs and is independent of the accretion rate of wind material onto the white dwarf (hence independent of Roche-lobe filling factor). Since these should be straightforward to identify, the complete lack of young, well detached strongly magnetic white dwarf binaries appears to be real.

\begin{figure}
  \begin{center}
    \includegraphics[width=\columnwidth]{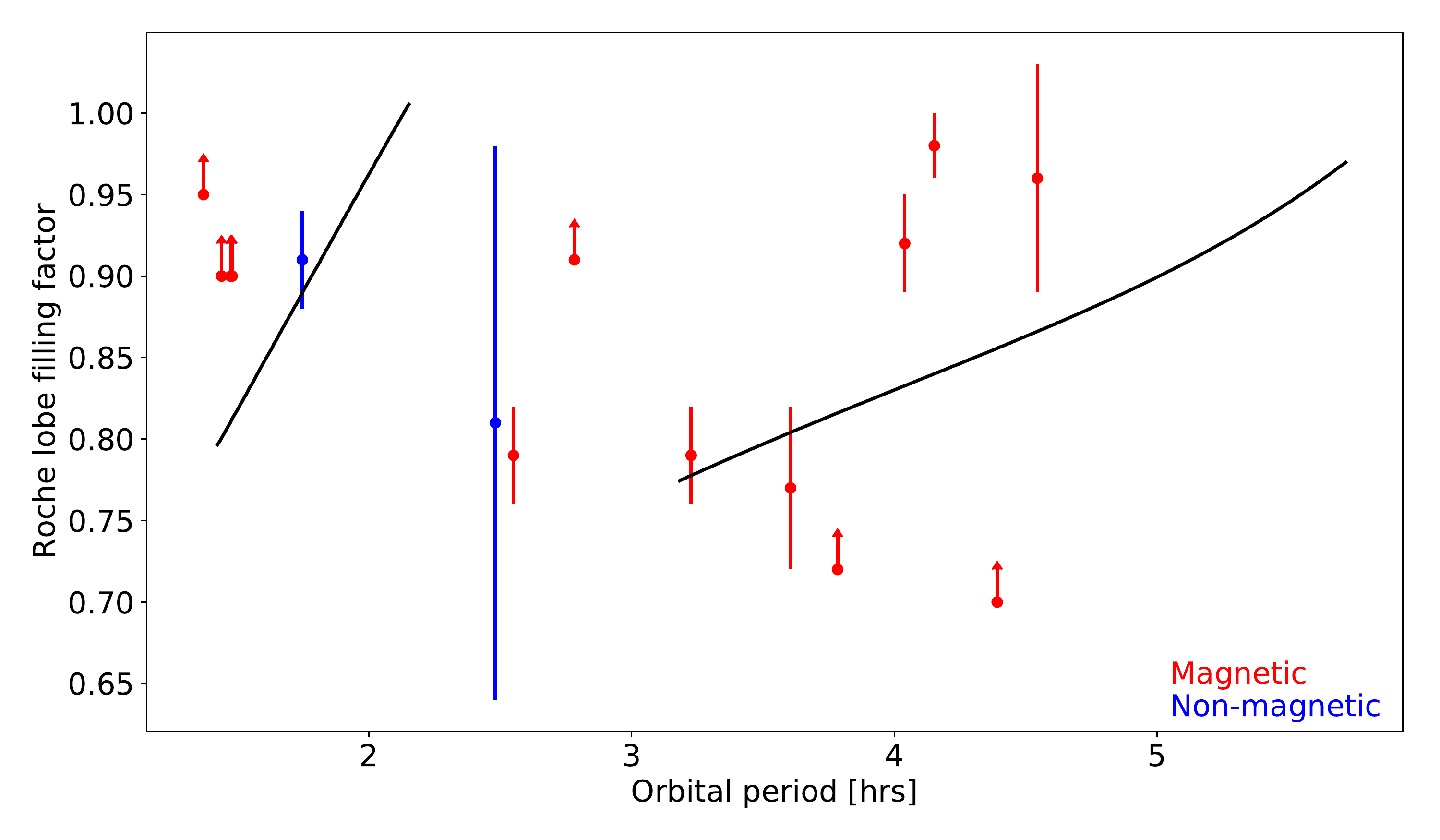}
    \caption{Roche-lobe filling factors for detached white dwarf plus main-sequence binaries as a function of their orbital periods. The black lines correspond to the filling factor that a previously Roche-lobe filling system would have if the donor relaxed to its equilibrium radius. Therefore, any systems above these lines could be detached cataclysmic variables.}
  \label{fig:max_detach}
  \end{center}
\end{figure}

Finally, we note that IL\,Leo stands out from the other magnetic white dwarf binaries analysed in this paper because of its short orbital period and low companion mass. However, IL\,Leo does share a number of similar properties to three other systems: SDSS\,J1212+0136 \citep{Burleigh06}, SDSS\,J1250+1549 and SDSS\,J1514+0744 \citep{Breedt12}. All of these systems contain magnetic white dwarfs and very low mass (or potentially sub-stellar) companions in binaries with periods very close to (but slightly above) the CV period minimum. The evolutionary status of these short period magnetic white dwarf binaries remains unclear \citep{Breedt12}. Like the longer period systems they may be pre-CV systems yet to come into contact or temporarily detached CVs, but there is also the possibility that these very short period magnetic white dwarf binaries are genuine CVs in which the donor star has become a brown dwarf and are now evolving towards longer periods, so-called "period-bounce" CVs. One example of a definite polar with a very low mass companion that entered such an extended low state is EF\,Eri (P$_\mathrm{orb}=81$\,min), which was an X-ray bright CV \citep{Charles79} until it switched off around 1997 \citep{Beuermann00} and has not been active since then. The accretion rates in period-bounce systems are low enough that they could appear similar to detached systems. There is some evidence from X-ray observations that SDSS\,J1212+0136 may be Roche-lobe filling \citep{Stelzer17}, which lends support to this interpretation. Therefore, these systems may form a distinct population from the longer period detached magnetic white dwarf binaries that have been the focus of this paper. However, IL\,Leo also differs from these systems in a number of ways. Its optical spectrum is dominated by strong cyclotron emission, unlike the other very short period systems, in which the photosphere of the white dwarf dominates. The white dwarf in IL\,Leo also has a stronger magnetic field and the orbital period of IL\,Leo is around 5 minutes shorter than these other systems. IL\,Leo may represent a slightly earlier phase in the evolution of these systems where the accretion rate is higher, generating stronger cyclotron emission. As the binary evolves towards longer periods the accretion rate will drop and in the future IL\,Leo may well resemble these other very short period magnetic white dwarf binaries.

\subsection{Identifying detached magnetic white dwarf binaries}

The small number of currently known magnetic white dwarfs in detached binaries limits the conclusions that we can draw from the population. Moreover, only one eclipsing system is currently known \citep{Parsons13_mag}. The discovery of new eclipsing systems would be particularly useful for distinguishing between the pre-CV and detached CV scenarios, since eclipsing systems allow a direct measurement of the radius of the main-sequence star (hence Roche-lobe filling factor) and can place tight constraints on the white dwarf mass, even though these systems are single lined. However, virtually all magnetic white dwarfs in detached binaries have so far been discovered serendipitously from their unusual optical spectra. As such, the sample is mostly limited to systems observed spectroscopically by SDSS.

Despite their rather unusual spectra, featuring strong cyclotron emission, detached magnetic white dwarf systems are not easily identifiable from their colours or position in the Hertzsprung-Russell Diagram. Figure~\ref{fig:HRD} shows the location of these binaries in the Hertzsprung-Russell Diagram relative to non-magnetic white dwarf plus main-sequence binaries. Unfortunately the magnetic systems are indistinguishable from the non-magnetic population, with both types of systems found between the white dwarf cooling track and the main-sequence track.

\begin{figure}
  \begin{center}
    \includegraphics[width=\columnwidth]{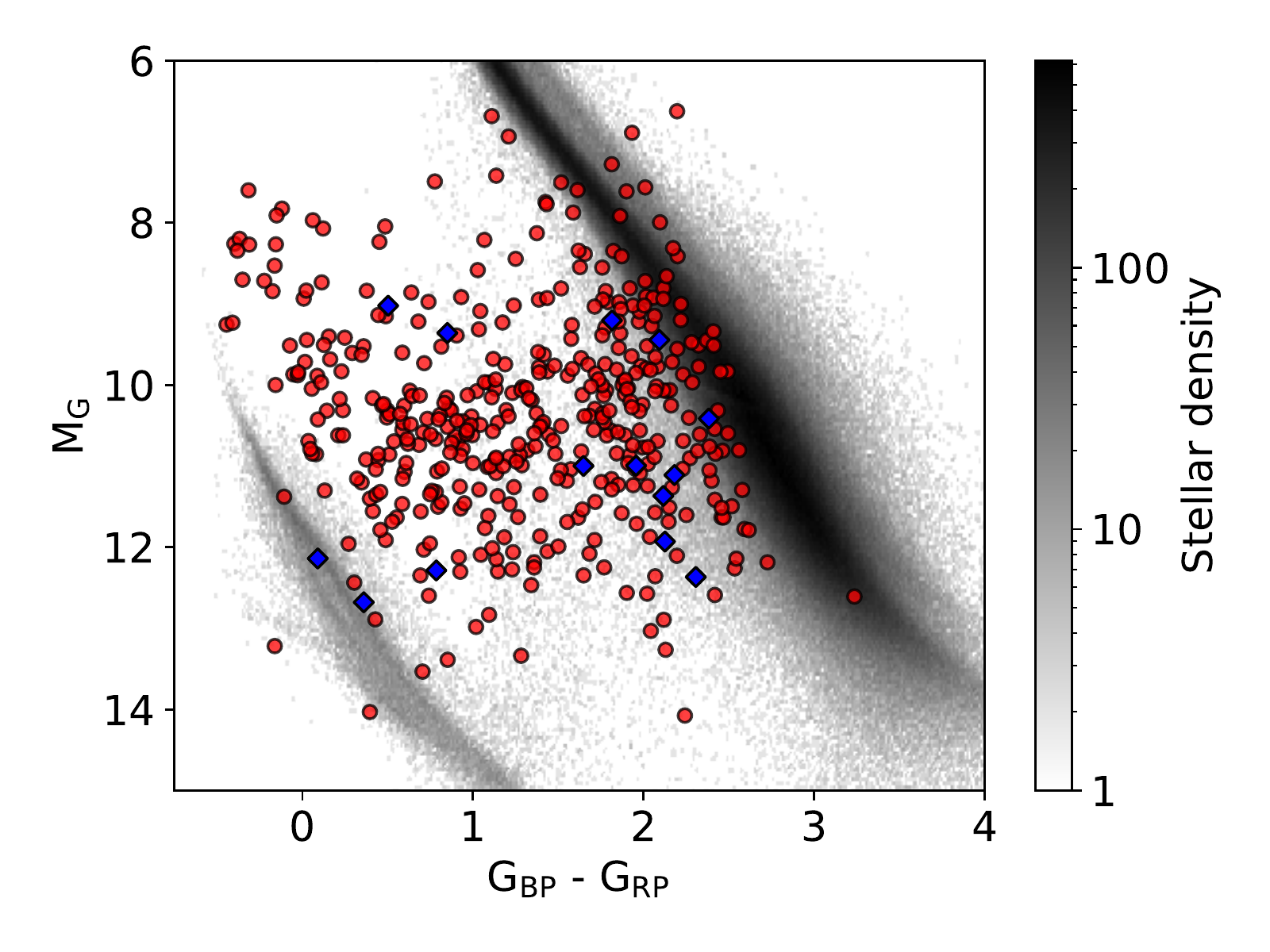}
    \caption{{\it Gaia} Hertzsprung-Russell Diagram showing the location of  detached magnetic (blue diamonds) and non-magnetic (red circles) white dwarf plus main-sequence binaries. A random sample of sources within 500\,pc of the Sun are shown in the background.}
  \label{fig:HRD}
  \end{center}
\end{figure}

\begin{figure}
  \begin{center}
    \includegraphics[width=\columnwidth]{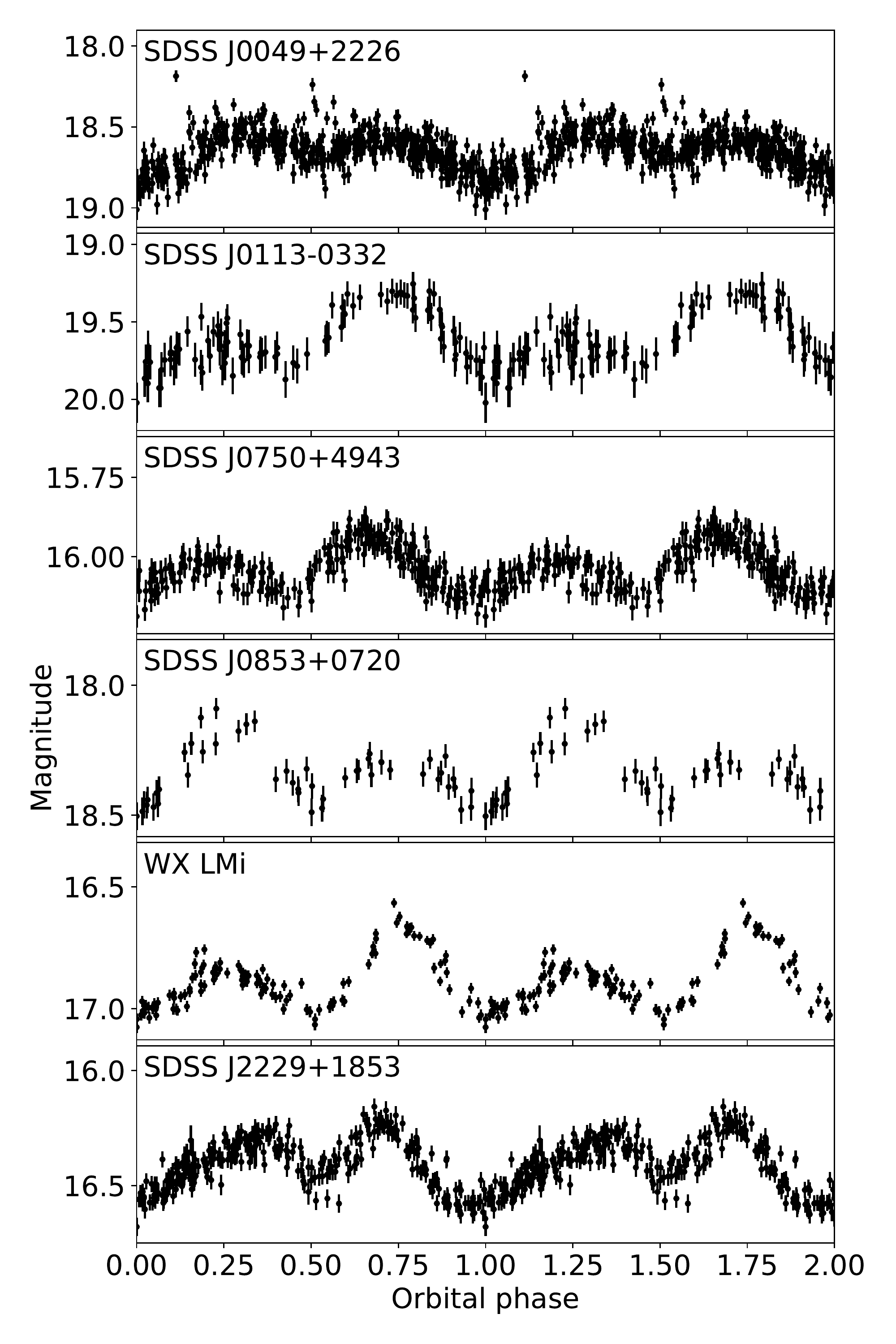}
    \caption{Phase-folded light curves of six detached magnetic white dwarf binaries displaying asymmetric light curves. Data for SDSS\,J0750+4943 and SDSS\,J2229+1853 are CRTS $V$ band data, all other light curves are ZTF $r$ band data. The asymmetric light curve shapes are a result of the magnetic pole(s) of the white dwarf rotating into view, on top of variations from the Roche-distorted main-sequence star.}
  \label{fig:CRTS}
  \end{center}
\end{figure}

Two systems in this study have demonstrated a useful method for identifying these systems via their light curves. SDSS\,J0750+4943 and SDSS\,J2229+1853 were both originally flagged as potential magnetic systems from their unusual asymmetric CRTS light curves (see Figure~\ref{fig:CRTS} and \citealt{Parsons15}). The double-peaked shape of the phase-folded light curve is reminiscent of ellipsoidal modulation, caused by the Roche-distorted main-sequence star presenting a different surface area throughout the orbit. However, in both cases one peak is substantially stronger than the other and the peaks have been shifted from the quadrature phases. This distortion of the light curve is due to the magnetic pole(s) of the synchronously rotating white dwarf moving into view, adding cyclotron emission to the light curve. This likely requires the system to be viewed from a relatively high inclination, in order for the magnetic pole(s) to disappear from view and generate the largest variations. Large scale multi-epoch photometric surveys such as ZTF and the Legacy Survey of Space and Time (LSST) are likely to identify many objects with similar looking light curves, offering us a chance to identify larger number of these types of systems without the need for prior spectroscopic observations. Four ZTF light curves of detached magnetic binaries (both confirmed and candidate systems) are also shown in Figure~\ref{fig:CRTS} along with two CRTS light curves demonstrating that these asymmetric light curves appear to be a common feature of these systems.

\section{Conclusions}

We have identified four new and seven candidate magnetic white dwarfs in detached binaries with main-sequence star companions. Using phase-resolved medium resolution spectroscopy we have analysed these four new systems in detail, along with two previously known magnetic white dwarf binaries and five non-magnetic DC white dwarf plus main-sequence star binaries. For each system we have constrained the stellar and binary parameters including, for the first time, the masses of the magnetic white dwarfs. We found that these magnetic white dwarfs are systematically more massive than non-magnetic white dwarfs in pre-cataclysmic binaries, but are consistent with the measured masses of white dwarfs in cataclysmic variables. We also found that, while some systems could be quite far from Roche-lobe filling, it is possible that all of these systems were Roche-lobe filling in the past, but the main-sequence stars are now closer to thermal equilibrium, although we cannot conclusively rule out the possibility that these systems are pre-cataclysmic binaries. Even if these systems are pre-cataclysmic binaries, they all have cooling ages of around a Gyr or more. Young detached magnetic white dwarf binaries remain elusive. Finally, we note that several of our newly identified magnetic white dwarf binaries have archival spectroscopy that show no clear evidence of magnetism, thus demonstrating that these systems can be missed when using spectroscopic data alone. We demonstrate that these magnetic white dwarf systems may be identifiable from their unusual light curves.

\section*{Data Availability Statement}

Raw and reduced X-shooter data are available through the ESO archive at \url{http://archive.eso.org/cms.html}. Raw IDS spectra are available through the Isaac Newton Group (ING) archive at \url{http://casu.ast.cam.ac.uk/casuadc/ingarch/query}, reduced IDS data are available on request.

\section*{Acknowledgements}

We thank the referee for their useful comments and suggestions. SGP acknowledges the support of a Science and Technology Facilities Council (STFC) Ernest Rutherford Fellowship. BTG was supported by the UK STFC grant ST/T000406/1 and by a Leverhulme Research Fellowship. TRM was supported by STFC grant ST/T000406/1. MRS thanks for support from FONDECYT (grant 1181404) and the millennium Nucleus for Planet Formation (NPF). The results presented in this paper are based on observations collected at the European Southern Observatory under programme IDs 099.D-0252 and 0101.D-0144.

\bibliographystyle{mnras}
\bibliography{magnetic_pcebs}

\appendix

\section{M dwarf velocity measurements}

\begin{table}
 \centering
  \caption{M dwarf velocity measurements presented in this paper}
  \label{tab:velocities}
  \begin{tabular}{@{}lcc@{}}
    \hline
    HJD(mid-exposure) & RV (\kms) & Err (\kms) \\
    \hline
    \multicolumn{2}{l}{\bf SDSS J022503.02+005456.2:} \\
2457996.8026337 &   71.2 & 3.2 \\
2457996.8073199 &   70.6 & 4.2 \\
2457996.8120076 &   71.5 & 3.8 \\
2457996.8166954 &   70.3 & 4.0 \\
2457996.8213834 &   75.7 & 6.1 \\
2457996.8260709 &   72.1 & 3.3 \\
2457996.8307592 &   71.2 & 3.3 \\
2457996.8354470 &   69.7 & 2.6 \\
2458012.7879503 &  -14.0 & 2.7 \\
2458012.7926343 &  -20.7 & 2.0 \\
2458012.7973319 &  -23.7 & 1.8 \\
2458012.8020175 &  -26.3 & 2.9 \\
2458012.8067054 &  -23.6 & 2.2 \\
2458012.8113943 &  -29.7 & 2.0 \\
2458012.8160809 &  -28.1 & 2.7 \\
2458012.8207624 &  -34.4 & 2.2 \\
	\multicolumn{2}{l}{\bf SDSS J075015.11+494333.2:} \\
2457065.3639952 &  195.2 & 6.6 \\
2457065.3746437 &  187.9 & 10.6 \\
2457065.3852192 &  119.4 & 7.0 \\
2457065.3957878 &   84.5 & 5.1 \\
2457065.4066982 &   18.2 & 4.9 \\
2457065.4172678 &  -38.4 & 4.9 \\
2457065.4292545 &  -80.9 & 4.4 \\
2457065.4417256 & -125.1 & 4.0 \\
2457065.4522908 & -124.3 & 4.0 \\
2457065.4638623 &  -95.5 & 3.8 \\
2457065.4749185 &  -69.6 & 3.5 \\
2457065.4870584 &    4.3 & 4.0 \\
2457065.4977579 &   54.3 & 4.0 \\
	\multicolumn{2}{l}{\bf SDSS J084841.17+232051.7:} \\
2457859.4829554 &   35.7 & 0.5 \\
2457859.4870521 &   23.4 & 0.5 \\
2457859.4911656 &   11.7 & 0.5 \\
2457859.4952740 &   -0.4 & 0.5 \\
2457859.4994626 &  -11.8 & 0.5 \\
2457859.5035729 &  -23.5 & 0.5 \\
2457859.5076811 &  -35.0 & 0.5 \\
2457859.5119777 &  -47.2 & 0.5 \\
2457859.5160924 &  -56.3 & 0.5 \\
2457859.5202022 &  -65.2 & 0.5 \\
2457859.5243221 &  -73.6 & 0.5 \\
2458224.5455325 &  -59.2 & 0.6 \\
2458224.5496477 &  -68.4 & 0.6 \\
2458224.5537677 &  -77.7 & 0.6 \\
2458224.5578703 &  -85.7 & 0.7 \\
2458224.5619978 &  -95.0 & 0.6 \\
2458224.5661160 & -102.1 & 0.6 \\
2458224.5702439 & -109.3 & 0.6 \\
2458224.5743663 & -116.5 & 0.6 \\
2458224.5784726 & -120.0 & 0.7 \\
2458226.5350704 &  -30.9 & 0.6 \\
2458226.5392074 &  -19.0 & 0.6 \\
2458226.5433374 &  -10.2 & 0.6 \\
2458226.5474558 &    2.1 & 0.6 \\
2458226.5515812 &   12.1 & 0.7 \\
2458226.5556985 &   23.5 & 0.7 \\
2458226.5598027 &   32.1 & 0.8 \\
2458226.5639267 &   44.7 & 0.7 \\
2458226.5680457 &   56.4 & 0.9 \\
2458227.5120304 &  -39.8 & 0.5 \\
2458227.5161571 &  -48.8 & 0.5 \\
    \hline
  \end{tabular}
\end{table}

\begin{table}
 \centering
  \contcaption{}
  \begin{tabular}{@{}lcc@{}}
    \hline
    HJD(mid-exposure) & RV (\kms) & Err (\kms) \\
    \hline
2458227.5202809 &  -58.6 & 0.5 \\
2458227.5244036 &  -67.7 & 0.5 \\
2458227.5285246 &  -77.5 & 0.5 \\
2458227.5326328 &  -88.5 & 0.6 \\
2458227.5367557 &  -93.0 & 0.5 \\
2458227.5408717 & -100.8 & 0.5 \\
2458227.5450031 & -108.2 & 0.5 \\
	\multicolumn{2}{l}{\bf SDSS J085336.03+072033.5:} \\
2457860.4815001 &  125.7 & 0.9 \\
2457860.4856083 &  153.1 & 0.9 \\
2457860.4897230 &  173.5 & 1.0 \\
2457860.4938327 &  190.4 & 1.0 \\
2457860.4979331 &  199.5 & 1.0 \\
2457860.5020421 &  204.5 & 0.9 \\
2457860.5061580 &  198.0 & 0.9 \\
2457860.5102652 &  184.4 & 1.0 \\
2457860.5143863 &  168.7 & 0.9 \\
2457860.5208586 &  131.6 & 0.9 \\
2457860.5249769 &   99.5 & 0.9 \\
2457860.5290831 &   65.7 & 0.9 \\
2457860.5331914 &   24.6 & 0.9 \\
2457860.5373009 &  -14.2 & 0.9 \\
2457860.5414082 &  -56.5 & 1.0 \\
2457860.5455213 &  -97.3 & 1.0 \\
2457860.5496251 & -135.5 & 1.0 \\
2457860.5537334 & -169.8 & 0.9 \\
	\multicolumn{2}{l}{\bf HS 0922+1333:} \\
2457859.5338685 &   18.9 & 0.5 \\
2457859.5379599 &   -1.9 & 0.5 \\
2457859.5420732 &  -19.6 & 0.5 \\
2457859.5461804 &  -38.1 & 0.5 \\
2457859.5503025 &  -56.3 & 0.5 \\
2457859.5544109 &  -71.4 & 0.5 \\
2457859.5585068 &  -85.1 & 0.5 \\
2457859.5626153 &  -94.7 & 0.5 \\
2457859.5667225 & -102.8 & 0.5 \\
2457859.5733788 & -105.7 & 0.5 \\
2457859.5774744 & -106.1 & 0.5 \\
2457859.5815834 & -102.2 & 0.5 \\
2457859.5856908 &  -95.5 & 0.5 \\
2457859.5898001 &  -86.2 & 0.5 \\
2457859.5939076 &  -75.6 & 0.5 \\
2457859.5980182 &  -62.4 & 0.5 \\
2457859.6021255 &  -46.4 & 0.5 \\
2457859.6062442 &  -29.7 & 0.5 \\
	\multicolumn{2}{l}{\bf SDSS J114030.06+154231.5:} \\
2457859.6166398 &   53.9 & 0.5 \\
2457859.6207427 &   53.7 & 0.5 \\
2457859.6248572 &   54.5 & 0.5 \\
2457859.6289670 &   55.0 & 0.5 \\
2457859.6330879 &   55.1 & 0.5 \\
2457859.6371966 &   55.1 & 0.5 \\
2457859.6413074 &   55.5 & 0.5 \\
2457859.6456340 &   55.9 & 0.5 \\
2457859.6497401 &   56.0 & 0.5 \\
2457859.6562043 &   58.7 & 0.5 \\
2457859.6603112 &   60.0 & 0.5 \\
2457859.6644279 &   59.4 & 0.5 \\
2457859.6685353 &   58.9 & 0.5 \\
2457859.6726448 &   60.0 & 0.5 \\
2457859.6767523 &   60.5 & 0.5 \\
2457859.6808483 &   61.5 & 0.5 \\
2457859.6849592 &   61.6 & 0.5 \\
    \hline
  \end{tabular}
\end{table}

\begin{table}
 \centering
  \contcaption{}
  \begin{tabular}{@{}lcc@{}}
    \hline
    HJD(mid-exposure) & RV (\kms) & Err (\kms) \\
    \hline
2457859.6890666 &   63.7 & 0.5 \\
2457860.6527829 &  -34.0 & 0.5 \\
2457860.6568901 &  -34.4 & 0.5 \\
2457860.6610006 &  -36.1 & 0.5 \\
2457860.6651079 &  -38.0 & 0.5 \\
2457860.6692269 &  -38.6 & 0.5 \\
2457860.6733380 &  -40.4 & 0.5 \\
2457860.6774386 &  -42.4 & 0.5 \\
2457860.6815422 &  -42.7 & 0.5 \\
2457860.6856520 &  -44.3 & 0.5 \\
2458254.5282508 &  -49.6 & 0.5 \\
2458254.5323637 &  -49.1 & 0.5 \\
2458254.5364850 &  -49.0 & 0.5 \\
2458254.5405903 &  -48.0 & 0.5 \\
2458254.5447134 &  -47.2 & 0.5 \\
2458254.5488322 &  -45.9 & 0.5 \\
2458254.5529627 &  -45.2 & 0.5 \\
2458254.5570851 &  -45.0 & 0.5 \\
2458254.5611942 &  -43.1 & 0.5 \\
	\multicolumn{2}{l}{\bf SDSS J131632.04-003758.0:} \\
2457859.6997413 &    9.7 & 0.6 \\
2457859.7038488 &   20.8 & 0.6 \\
2457859.7079576 &   32.9 & 0.6 \\
2457859.7120561 &   45.5 & 0.6 \\
2457859.7161644 &   56.3 & 0.6 \\
2457859.7202732 &   66.9 & 0.5 \\
2457859.7243900 &   76.8 & 0.6 \\
2457859.7285012 &   86.4 & 0.6 \\
2457859.7326015 &   97.0 & 0.6 \\
	\multicolumn{2}{l}{\bf SDSS J145238.12+204511.9:} \\
2457859.7397896 & -398.8 & 1.2 \\
2457859.7438907 & -396.9 & 1.2 \\
2457859.7480076 & -365.3 & 1.3 \\
2457859.7521164 & -319.5 & 1.2 \\
2457859.7562253 & -254.9 & 1.3 \\
2457859.7603342 & -179.0 & 1.2 \\
2457859.7644433 &  -93.6 & 1.3 \\
2457859.7685519 &   -8.9 & 1.3 \\
2457859.7726606 &   77.4 & 1.2 \\
2457859.7790416 &  197.4 & 1.1 \\
2457859.7831448 &  257.7 & 1.2 \\
2457859.7872438 &  298.6 & 1.2 \\
2457859.7913531 &  319.6 & 1.2 \\
2457859.7954723 &  322.0 & 1.2 \\
2457859.7995928 &  298.8 & 1.1 \\
2457859.8037007 &  259.7 & 1.2 \\
2457859.8078105 &  200.8 & 1.2 \\
2457859.8119183 &  131.8 & 1.3 \\
	\multicolumn{2}{l}{\bf SDSS J220848.32+003704.6:} \\
2457995.5544827 &   72.3 & 2.5 \\
2457995.5591719 &  131.3 & 2.5 \\
2457995.5638697 &  185.2 & 2.1 \\
2457995.5685470 &  220.8 & 2.0 \\
2457995.5732435 &  251.9 & 2.7 \\
2457995.5779321 &  243.1 & 1.9 \\
2457995.5826196 &  225.5 & 2.0 \\
2457995.5873061 &  194.5 & 2.6 \\
2458000.6254337 &  142.5 & 3.3 \\
2458000.6301217 &  205.8 & 3.3 \\
2458000.6348104 &  232.0 & 2.9 \\
2458000.6394954 &  240.4 & 2.4 \\
2458000.6488727 &  219.7 & 2.2 \\
2458000.6441842 &  248.8 & 2.3 \\
    \hline
  \end{tabular}
\end{table}

\begin{table}
 \centering
  \contcaption{}
  \begin{tabular}{@{}lcc@{}}
    \hline
    HJD(mid-exposure) & RV (\kms) & Err (\kms) \\
    \hline
2458000.6535578 &  188.9 & 2.1 \\
2458000.6582466 &  137.7 & 2.1 \\
	\multicolumn{2}{l}{\bf SDSS J222918.95+185340.2:} \\
2457995.6374760 &  136.8 & 0.8 \\
2457995.6415812 &  116.6 & 0.9 \\
2457995.6456889 &   88.4 & 1.1 \\
2457995.6498000 &   58.9 & 0.9 \\
2457995.6539078 &   31.1 & 0.9 \\
2457995.6580156 &    1.8 & 1.0 \\
2457995.6621389 &  -28.0 & 0.9 \\
2457995.6662474 &  -55.3 & 0.9 \\
2457995.6703563 &  -80.2 & 0.9 \\
2457995.6771421 & -116.8 & 0.8 \\
2457995.6812423 & -136.9 & 0.8 \\
2457995.6853550 & -154.9 & 0.8 \\
2457995.6894628 & -166.9 & 0.8 \\
2457995.6935752 & -178.3 & 0.8 \\
2457995.6976816 & -182.1 & 0.9 \\
2457995.7017892 & -189.8 & 0.8 \\
2457995.7058996 & -188.7 & 0.9 \\
2457995.7100199 & -183.5 & 0.8 \\
2457996.6735976 & -133.3 & 0.8 \\
2457996.6777038 & -116.0 & 0.9 \\
2457996.6818124 &  -90.8 & 0.9 \\
2457996.6859214 &  -64.8 & 0.9 \\
2457996.6900201 &  -40.1 & 0.9 \\
2457996.6941275 &  -13.4 & 0.8 \\
2457996.6982373 &   15.8 & 0.8 \\
2457996.7023461 &   40.9 & 0.7 \\
2457996.7064656 &   68.6 & 0.8 \\
2457996.7131085 &  110.7 & 0.8 \\
2457996.7172065 &  132.5 & 0.8 \\
2457996.7213169 &  156.5 & 0.8 \\
2457996.7254258 &  172.6 & 0.8 \\
2457996.7295491 &  189.5 & 0.9 \\
2457996.7336534 &  198.8 & 0.9 \\
2457996.7377623 &  205.8 & 0.9 \\
2457996.7418711 &  213.9 & 0.9 \\
2457996.7459798 &  213.8 & 0.9 \\
    \hline
  \end{tabular}
\end{table}
    
\section{White dwarf velocity measurements}

\begin{table}
 \centering
  \caption{White dwarf velocity measurements for SDSS J084841.17+232051.7 from the Ca\,{\sc ii} 3934{\AA} absorption line}
  \label{tab:wd_velocities}
  \begin{tabular}{@{}lcc@{}}
    \hline
    HJD(mid-exposure) & RV (\kms) & Err (\kms) \\
    \hline
2457859.4846316 &  46.5 & 4.5 \\
2457859.4919063 &  53.8 & 4.9 \\
2457859.4991749 &  68.6 & 4.9 \\
2457859.5064655 &  88.9 & 4.9 \\
2457859.5137851 & 109.6 & 4.9 \\
2457859.5210477 & 131.8 & 5.1 \\
2458224.5472608 & 119.8 & 4.9 \\
2458224.5546431 & 140.2 & 4.5 \\
2458224.5620364 & 150.8 & 4.6 \\
2458224.5694316 & 160.8 & 4.7 \\
2458224.5768256 & 174.5 & 4.9 \\
2458226.5367985 &  84.6 & 4.7 \\
2458226.5441880 &  74.3 & 4.6 \\
2458226.5515741 &  51.4 & 4.9 \\
2458226.5589551 &  34.5 & 4.8 \\
2458226.5663385 &  13.4 & 5.1 \\
2458227.5137585 & 107.5 & 4.9 \\
2458227.5211569 & 125.4 & 5.0 \\
2458227.5285455 & 140.4 & 5.1 \\
2458227.5359294 & 149.5 & 5.0 \\
2458227.5433194 & 168.6 & 4.7 \\
    \hline
  \end{tabular}
\end{table}

\bsp
\label{lastpage}
\end{document}